\documentclass[final,5p,times,nofootinbib,twocolumn]{elsarticle}

\usepackage{amssymb}
\usepackage{mathtools}
\usepackage{amsmath}    
\usepackage{amsfonts} 
\usepackage{latexsym}
\usepackage{graphics}
\usepackage[english]{babel}
\usepackage{graphicx}
\usepackage{epsfig}
\usepackage{layout}
\usepackage{pgfplots}
\usepackage{palatino}
\usepackage{pgfplotstable}
\usepackage{tikz}
\usepackage{braket}
\usepackage{mathpazo}
\usepackage{bbding}
\usepackage{wasysym}
\DeclareMathOperator{\sech}{sech}
\newcommand{\ii }{\'{\i}}
\usepackage[utf8]{inputenc}
\usepackage[T1]{fontenc}
\usepackage{lmodern} 
\usepackage{listings}
\usepackage{color}
\usepgfplotslibrary{fillbetween}
\usetikzlibrary{arrows}
\selectcolormodel{rgb}
\definecolor{Mygreen}{rgb}{0.15,0.75,0.45}
\usetikzlibrary{patterns}
\pgfplotsset{compat=newest}
\usepackage{pict2e}
\usepackage{graphicx}
\usepackage[compact]{titlesec}
\titlespacing{\section}{0pt}{2ex}{2ex}
\renewcommand{\baselinestretch}{1.0}
\selectcolormodel{cmyk}
\journal{Journal of Magnetism and Magnetic Materials}

\begin{document}

\begin{frontmatter}
\ead{hvivasc@unal.edu.co}
\title{Anomalous Magnetism for Dirac Electrons in Two Dimensional Rashba Systems}



%
%
\author{H. Vivas C.}
\address{Departamento de F\'{i}sica,
Universidad Nacional de Colombia, Sede Manizales, A.A. 127, Colombia}
\begin{abstract}
Spin-spin correlation function response in the low electronic density regime and externally applied electric field is evaluated for 2D metallic crystals under Rashba-type coupling, fixed number of particles and two-fold energy band structure. Intrinsic Zeeman-like effect on electron spin polarization, density of states, Fermi surface topology and transverse magnetic susceptibility are analyzed in the zero temperature limit. A possible magnetic state for Dirac electrons depending on the zero field band gap magnitude under this conditions is found.
\end{abstract}

\end{frontmatter}


\section{Introduction}
Subtle relativistic effects on reduced dimensional electronic systems have brought exciting perspectives on fundamental physics and technological advances since the Rashba's breakthrough \cite{Rashba,SRFP,B,B2}. Spin Orbit Interaction dwells in the always evolving spintronics world, providing interesting applications based on its subsequent, wide and sophisticated phenomena. For instance, magnetic switching control via induced current on metal/ferromagnet/oxide trilayers at room temperature \cite{Perez}, tunable spin-orbit strength via stoichiometry manipulation on deposited concentration of Bi atoms on Bi$_{x}$Pb$_{1-x}$/Ag alloys \cite{control}, quantized Hall conductance on doped Bi$_2$Te$_{3}$ layered arrays without external magnetic field \cite{Hall}, characteristic Knight shift behavior in non-centrosymmetric superconducting CrIrSe$_3$ crystals below critical temperature \cite{kn}, or Rashba interaction control for out-of-plane Zeeman spin polarization on transition metals such as WSe$_2$, MoS$_2$ via biased voltage \cite{Yuan}, constitute few examples that demonstrate the currently hectic activity in this area. In this paper, we discuss the zeroth order transverse spin-spin susceptibility response for 2D Dirac interacting electrons in the low density regime and zero temperature, under an externally applied electric field on the plane. We derive general expressions at finite temperature for the correlations functions on arbitrary Dirac spin directions, as well as the features of the density of states (DOS) in the limit of the Fermi energy instability and fixed number of particles. 
\section{Spin-Spin Correlation Function Formalism}
The Hamiltonian formulation for 2D magnetically polarized surfaces or interfaces in a non-interacting electron system under an externally applied electric field $\mathbf{E}$ might be approached by taking the lowest-order coupling between the electron momentum in the $XY$ plane $\mathbf{k}=(k_{x},k_{y},0)$,  its spin $\hat{\vec{\sigma}}$ and $\mathbf{E}$: $\hat{\mathcal{H}}_{\mathbf{k}}\sim\hat{K}_{\mathbf{k}}+\hat{\vec{\sigma}}\cdot{\mathbf{(k\times E)}}$, where $\hat{K}_{\mathbf{k}}=(\hbar^{2}k^{2}/2m^{\star})\hat{\mathbf{1}}$ corresponds to the single free-particle kinetic energy operator \cite{Zhang,Winkler,Barnes,Qi}. For an arbitrary field orientation, $\hat{\mathcal{H}}_{\mathbf{k}}\sim\hat{K}_{\mathbf{k}}+\hat{\sigma}_{Z}(k_{x}E_{y}-k_{y}E_{x})+(\hat{\sigma}_{X}k_{y}-\hat{\sigma}_{Y}k_{x})E_{z}$, where the last term is recognized as the typical Rashba-type interaction. By introducing the appropriate constants, the upper-lower ($\pm$) double band structure for Dirac electrons might be taken as $\hat{\mathcal{H}}_{\mathbf{k}}^{\pm}=\hat{K}_{\mathbf{k}}+\hat{\sigma}_{Z}\Delta_{\mathbf{k}}-\alpha\hat{\sigma}_{Y}k_{x}\pm\alpha\hat{\sigma}_{X}k_{y}$. Defining the effective magnetic field $\gamma\mathbf{B_{\Sigma}}=(-\alpha k_{y},\alpha k_{x},-\Delta_{\mathbf{k}})$, the complete electron Hamiltonian takes the form:
\begin{equation}\label{e00}
\hat{\mathcal{H}}_{\mathbf{k}}^{\pm}=\hat{K}_{\mathbf{k}}-\gamma\mathbf{\hat{S}\cdot B_{\Sigma}},
\end{equation}
with the spin basis $\mathbf{\hat{S}}\equiv\left(\hat{\sigma}_{Z}\otimes\hat{\sigma}_{X},I\otimes\hat{\sigma}_{Y},I\otimes\hat{\sigma}_{Z}\right)$ and $I$, $\hat{\sigma}_{j}$ as the $2\times 2$ identity and Pauli matrices respectively. Operators $\hat{S}_{j}$ satisfy the necessary anticommutation rules $[\hat{S}_{i},\hat{S}_{j}]_{+}=2\delta_{ij}$. The Zeeman-like term $\mathcal{\hat{H}}_{\mathbf{\Sigma}}=-\gamma\mathbf{\hat{S}\cdot B_{\Sigma}}$ in Eq. (\ref{e00}) has a Dirac-type form, and its genesis can be explained, among several approaches, from fairly simple geometric-based arguments for materials with inversion symmetry \cite{Tan,MM,Rodin}. Specifically, the 2D Dirac equation $\mathcal{\hat{H}}_{\mathbf{\Sigma}}=\alpha\hat{\gamma}^{0}(\hat{\vec{\gamma}}\cdot\mathbf{k}+\alpha m^{\star})$ reduces into the Zeeman-like Hamiltonian straightforwardly under the set of transformations $\alpha^{2}m^{\star}\equiv\Delta_{\mathbf{k}}=\Delta_{0}+(k_{x}\bar{E}_{y}-k_{y}\bar{E}_{x})$, where $\Delta_{0}$ corresponds to the energy band gap magnitude at zero field. The parameter $\alpha$ denotes the typical Rashba spin-orbit interaction constant, although contributions due to crystal asymmetries might be taken into account via Dresselhaus Hamiltonian \cite{Chap}. Upon this representation, the Dirac matrices $\hat{\gamma}^{\mu}=(\hat{\gamma}^{0},\hat{\vec{\gamma}})$, $\hat{\gamma}^{\mu}=(I\otimes\hat{\sigma}_{Z},iI\otimes\hat{\sigma}_{X},i\hat{\sigma}_{Z}\otimes\hat{\sigma}_{Y})$ must fulfill the constraint $[\hat{\gamma}^{\mu},\hat{\gamma}^{\nu}]_{+}=2g^{\mu\nu}=2\mbox{diag}(1,-1,-1)$. In a matrix-block scheme:
\begin{equation}\label{e1}
\mathcal{\hat{H}}_{\mathbf{\Sigma}}=
\begin{pmatrix}
A_{\mathbf{k}+}&0\\
0&A_{\mathbf{k}-}\\
\end{pmatrix},
\end{equation}
where $A_{\mathbf{k}\sigma}\equiv A_{\mathbf{k}\pm}$ is giving by:
\begin{equation}
A_{\mathbf{k}\sigma}=
\begin{pmatrix}
\Delta_{\mathbf{k}}&i\alpha ke^{-i\sigma\phi_{\mathbf{k}}}\\
-i\alpha ke^{i\sigma\phi_{\mathbf{k}}}&-\Delta_{\mathbf{k}}\\
\end{pmatrix},
\end{equation}
and $\tan{\phi_{\mathbf{k}}}=k_{y}/k_{x}$. Eigenvalues of $\mathcal{\hat{H}_{\mathbf{k}}}^{\sigma}$ provides the two-fold energy spectrum:
\begin{equation}\label{energy0}
\varepsilon_{\mathbf{k}\sigma}^{0}=\frac{\hbar^{2}k^2}{2m^{\star}}+\sigma\sqrt{\alpha^{2}k^{2}+\Delta_{\mathbf{k}}^{2}},
\end{equation}
where $\sigma\equiv\pm 1$ labels the band index. The finite temperature Green's propagator associated to $\mathcal{\hat{H}}_{\mathbf{k}}^{\sigma}$ is calculated from its orthonormalized eigenvectors  $\ket{u_{\sigma\mathbf{k}\uparrow}}=\ket{i\sigma F_{\mathbf{k}\sigma}e^{-i\phi_{\mathbf{k}}},1,0,0}/(1+F^{2}_{\mathbf{k}\sigma})^{1/2}$,  $\ket{u_{\sigma\mathbf{k}\downarrow}}=\ket{0,0,i\sigma F_{\mathbf{k}\sigma}e^{i\phi_{\mathbf{k}}},1}/(1+F^{2}_{\mathbf{k}\sigma})^{1/2}$ \cite{Fetter}:
\begin{equation}\label{G0}
\mathcal{G}_{ij}^{0}\left(\mathbf{k},i\omega_{n}\right)=\sum_{\sigma=\lbrace\pm\rbrace}\frac{M^{\sigma}_{ij}\left(\mathbf{k}\right)}{i\omega_{n}-\hbar^{-1}(\varepsilon_{\mathbf{k}\sigma}^{0}-\mu)},
\end{equation}
with $\omega_{n}$ as the (Fermionic) Matsubara frequencies, $\mu$ is the chemical potential and the matrix elements $\mathbf{M}^{\sigma}\left(\mathbf{k}\right)\equiv\sum_{s=\uparrow, \downarrow}\ket{u_{\sigma\mathbf{k}s}}\bra{u_{\sigma\mathbf{k}s}}$ are defined through $(1+F^{2}_{\mathbf{k}\sigma})\times M^{\sigma}_{ij}\left(\mathbf{k}\right)=$
\begin{equation}
\begin{pmatrix}
  F^{2}_{\mathbf{k}\sigma}& i\sigma F_{\mathbf{k}\sigma}e^{-i\phi_{\mathbf{k}}}&0&0 \\
-i\sigma F_{\mathbf{k}\sigma}e^{i\phi_{\mathbf{k}}}&1&0&0\\
0&0&F^{2}_{\mathbf{k}\sigma}&i\sigma F_{\mathbf{k}\sigma}e^{i\phi_{\mathbf{k}}} \\
0&0&-i\sigma F_{\mathbf{k}\sigma}e^{-i\phi_{\mathbf{k}}}&1\\
 \end{pmatrix},
\end{equation}
with $\alpha kF_{\mathbf{k}\sigma}=\sigma\Delta_{\mathbf{k}}+(\alpha^{2}k^{2}+\Delta_{\mathbf{k}}^2)^{1/2}$.
 The average value for the spin operator $\mathbf{\hat{S}}$ (per unit of surface $\mathcal{S}$) is calculated from the prescription:
\begin{equation}\label{szz1}
\frac{\langle\mathbf{\hat{S}}\rangle}{\mathcal{S}}=\frac{1}{\hbar\beta}\sum_{n,\mathbf{k}}\mbox{Tr}\lbrace\mathbf{\hat{S}}\mathcal{G}^{0}(\mathbf{k},i\omega_{n})\rbrace,
\end{equation}
where Tr corresponds to the trace operator. Direct calculation for the $Z$-compound leads into:
\begin{equation}\label{no}
\frac{\langle\hat{S}_{Z}\rangle}{\mathcal{S}}=-2\sum_{\mathbf{k}}\frac{\Delta_{\mathbf{k}}}{\varepsilon^{g}_{\mathbf{k}}}\left[\frac{\sinh{(\beta\varepsilon^{g}_{\mathbf{k}})}}{\cosh{(\beta\bar{\mu}_{\mathbf{k}})}+\cosh{(\beta\varepsilon^{g}_{\mathbf{k}})}}\right],
\end{equation}
with $\beta^{-1}=k_{B}T$, $\varepsilon^{g}_{\mathbf{k}}=(\alpha^{2}k^2+\Delta_{\mathbf{k}}^2)^{1/2}$ and $\bar{\mu}_{\mathbf{k}}=\mu-\hbar^{2}k^{2}/2m^{\star}$. Energy gap structure is affected by $\langle \hat{S}_{Z}\rangle$ under the minimum spin-effective interaction term $\mathcal{\hat{H}}_{I}=-J\langle \hat{S}_{Z}\rangle \hat{S}_{Z}$ or in equivalent form, by the transformation $\Delta_{\mathbf{k}}\rightarrow\Delta_{\mathbf{k}}-J\langle\hat{S}_{Z}\rangle$. Integration of Eq. (\ref{no}) at zero temperature limit, zero applied field, non interacting spins ($J=0$) and low density regime provides the exact result:
\begin{equation}
\langle\hat{S}_{Z}\left(0\right)\rangle=-\frac{2k_{0}^{2}\mathcal{S}\bar{\Delta}_{0}}{\pi}\sqrt{1+2\bar{\mu}+\bar{\Delta}_{0}^{2}}.
\end{equation}
The existence of this particular state of magnetic polarization is biased by the value of the Fermi energy for the range $\bar{\mu}\geq-(1+\bar{\Delta}_{0}^{2})/2$ with $\bar{\mu}=\mu/2E_{0}$, $\bar{\Delta}_{0}=\Delta_{0}/2E_{0}$, $k_{0}=m^{\star}\alpha/\hbar^{2}$, $2E_{0}=\alpha k_{0}$. For gapless systems in the regime $J\neq 0$ the average spin orientation on $Z$ direction is still feasible for $\bar{\mu}\leq (\zeta^{2}-1)/2$, $\zeta=\pi/2\bar{J}k_{0}^{2}\mathcal{S}$: 
\begin{equation}
 \langle\hat{S}_{Z}(0)\rangle=\frac{1}{\bar{J}}\sqrt{\zeta^2-2\bar{\mu}-1},
\end{equation}
with $\bar{J}=J/2E_{0}$.
Two mechanisms for possible magnetization in out-of-plane direction at zero temperature and without external field are unveiled: both are intrinsically induced by the presence of the component of an effective Zeeman field due either to the band gap $B_{\Sigma Z}\sim-\Delta_{0}/\gamma$ for $J=0$, or the non spin-spin interaction and $B_{\Sigma Z}\sim -J\langle\hat{S}_{Z}\rangle/\gamma$ for $\Delta_{0}=0$. The model also suggests that, for the last case, there is a resulting magnetization in the range $\bar{\mu}<0$ generated by the coupling parameter $J$ only if the size of the sample follows the restriction $\mathcal{S}\geq \pi\hbar^{2}/2m^{\star}J$.
The carrier density is calculated from $N/\mathcal{S}=(\hbar\beta)^{-1}\sum_{n,\mathbf{k}}\mbox{Tr}\lbrace\mathcal{G}^{0}(\mathbf{k},i\omega_{n})\rbrace$ and in the low density regime ($\bar{\mu}<0$) the number of particles is obtained by integrating $N/\mathcal{S}$ upon appropriate limits $\bar{k}_{F\pm}=(2(1+\bar{\mu}\pm(1+2\bar{\mu}+\bar{\Delta}^{2}_{0})^{1/2}))^{1/2}$, $k_{F\sigma}=k_{0}\bar{k}_{F\sigma}$, with the result: $N_{-}/\mathcal{S}=(2k_{0}^{2}/\pi)(1+2\bar{\mu}+\bar{\Delta}_{0}^{2})^{1/2}$. Bare susceptibility $\chi^{0}$ can also be written in terms of the spin-spin correlation function $\mathcal{D}^{0}$ by setting $\chi^{0}_{ij}\left(q\right)=-\mu_{0}^{2}\mathcal{D}_{ij}^{0}\left(q\right)$, and 
\begin{equation}\label{dij}
\mathcal{D}^{0}_{ij}\left(q\right)=\frac{2}{\mathcal{S}}\frac{1}{\beta}\sum_{n,\mathbf{k}}\mbox{Tr}\lbrace\mathcal{G}^{0}\left(k\right)\hat{S}_{i}\mathcal{G}^{0}\left(k+q\right)\hat{S}_{j}\rbrace ,
\end{equation}
with the reduced notation $k\equiv\left(\mathbf{k},i\omega_{n}\right)$, and $q\equiv\left(\mathbf{q},i\nu_{m}\right)$ \cite{Yan}. 
Wavevector $\mathbf{q}$ might be interpreted as the exchanging momentum for (Dirac) spin-spin collective excitations. Similar developments have been performed for describing the plasmon dispersion relationships on helical liquid state in Bi$_{2}$Se$_3$ \cite{SP} and intrinsic graphene layers \cite{Wang}. Equation (\ref{dij}) may be expressed in terms of the generalized products:
\begin{equation}
\begin{multlined}
\frac{1}{\beta}\sum_{n,\mathbf{k}}\mathcal{G}_{ij}^{0}\left(k\right)\mathcal{G}_{kl}^{0}\left(k+q\right)=
\\
\sum_{\mathbf{k}}\sum_{\sigma\sigma^{\prime}}M_{ij}^{\sigma}(\mathbf{k})M_{kl}^{\sigma^{\prime}}(\mathbf{k+q})\Pi_{\sigma\sigma^{\prime}}^{0}\left(i\nu_{m},\mathbf{q},\mathbf{k}\right),
\end{multlined}
\end{equation}
where
\begin{equation}
\Pi_{\sigma\sigma^{\prime}}^{0}\left(i\nu_{m},\mathbf{q},\mathbf{k}\right)=-\frac{n_{\mathbf{k+q}\sigma^{\prime}}^{0}-n_{\mathbf{k}\sigma}^{0}}{i\hbar\nu_{m}-(\varepsilon_{\mathbf{k+q}\sigma^{\prime}}^{0}-\varepsilon_{\mathbf{k}\sigma}^{0})},
\end{equation}
with $n_{\mathbf{k}\sigma}^{0}=(1+\exp{[\beta(\varepsilon_{\mathbf{k}\sigma}^{0}-\mu)]})^{-1}$. In the \emph{static} case, $\nu_{m}\rightarrow 0$, with $\mathbf{q\rightarrow 0}$,  the term $\mathcal{D}^{0}_{ij}\left(0\right)$ takes the form:
\begin{equation}\label{DZZ}
\mathcal{D}^{0}_{ij}\left(0\right)=-\frac{\beta}{\mathcal{S}}\sum_{\mathbf{k},\sigma}B^{ij}_{\mathbf{k}}\sech^{2}{\left(\frac{\beta X_{\mathbf{k}\sigma}}{2}\right)}
\end{equation}
\begin{equation*}
-\frac{4}{\mathcal{S}}\sum_{\mathbf{k}}\frac{C^{ij}_{\mathbf{k}}}{\left(X_{\mathbf{k}+}-X_{\mathbf{k}-}\right)}\left[\tanh\left(\frac{\beta X_{\mathbf{k}+}}{2}\right)-\tanh\left(\frac{\beta X_{\mathbf{k}-}}{2}\right)\right],
\end{equation*}
with $B^{ij}_{\mathbf{k}}$, $C^{ij}_{\mathbf{k}}$ and $X_{\mathbf{k}\sigma}$ defined in \cite{coef}. Bare susceptibility can be calculated from (\ref{DZZ}) with the particular choice $\hat{S}_{i}=\hat{S}_{j}\equiv\mathbf{\hat{1}}$ in Eq. (\ref{dij}), or under the conditions $B_{\mathbf{k}}^{ij}=1$, $C_{\mathbf{k}}^{ij}=0$, leading into $\chi^{0}=\mu_{0}^{2}\beta/\mathcal{S}\sum_{\mathbf{k},\sigma}\sech^{2}(\beta X_{\mathbf{k}\sigma}/2)$, whose zero temperature limit converges to the well known result $\chi^{0}=4\mu_{0}^{2}/\mathcal{S}\sum_{\mathbf{k},\sigma}\delta(\varepsilon_{\mathbf{k}\sigma}^{0}-\mu)$.
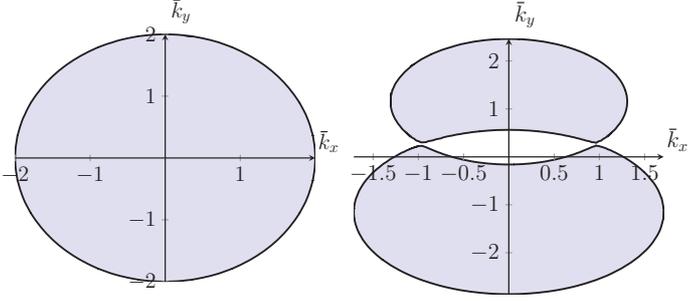
\begin{figure}
\begin{minipage}[t]{4.275cm}
     \centering
        \resizebox{\linewidth}{!}{%
\pgfplotsset{every axis/.append style={font=\Large,
extra description/.code={5
}}}
\begin{tikzpicture}
\begin{axis}[axis lines = middle,
    thick,no markers,
xlabel = {\makebox[0pt]{\hspace{25pt}\Large $\bar{k}_{x}$}},
   ylabel = {\raisebox{15pt}[0pt][0pt]{\Large $\bar{k}_{y}$}},
axis on top]
\addplot[black,smooth,very thick,color=black,fill=blue!10]
coordinates{(-0.8957459538775328, -1.7886030967346758) (-0.8928571428571426, 
-1.7899411655300907) (-0.8845745472826915, -1.7939968812887368) 
(-0.7723096977773807, -1.8437382692059525) (-0.714285714285714, 
-1.868475985585385) (-0.6966831378855314, -1.875) 
(-0.6431815687113001, -1.8931815687113003) (-0.5894087524043236, 
-1.9105912475956761) (-0.5357142857142854, -1.9248695999462928) 
(-0.5055334733923842, -1.934104901963813) (-0.3753810253482673, 
-1.9642857142857142) (-0.3604706770061781, -1.9676135341490355) 
(-0.3571428571428568, -1.9681330321464447) (-0.352381473652322, 
-1.969047097776249) (-0.2678571428571425, -1.9819683021148051) 
(-0.20326585791427762, -1.9889801436285635) 
(-0.17857142857142824, -1.9906563620406859) (-0.1490683017697082, 
-1.9937888410874343) (-0.0345987937225406, -1.9988845080082551) 
(0.0001, -1.9982072874278278) 
(0.034598793722541266, -1.9988845080082551) (0.14906830176970903, 
-1.993788841087434) (0.1785714285714289, -1.9906563620406856) 
(0.20326585791427815, -1.9889801436285635) (0.3523814736523229, 
-1.9690470977762489) (0.3571428571428575, -1.9681330321464445) 
(0.3604706770061786, -1.9676135341490353) (0.3753810253482671, 
-1.9642857142857142) (0.5055334733923849, -1.934104901963813) 
(0.535714285714286, -1.9248695999462928) (0.5894087524043247, 
-1.9105912475956757) (0.6431815687113004, -1.8931815687113) 
(0.6966831378855317, -1.875) (0.7142857142857146, 
-1.8684759855853847) (0.7723096977773811, -1.843738269205952) 
(0.8845745472826934, -1.7939968812887355) (0.8928571428571432, 
-1.78994116553009) (0.895745953877533, -1.7886030967346753) 
(0.9010601002563274, -1.7857142857142856) (1.010476973039604, 
-1.7247626873253177) (1.0714285714285718, -1.685604518080289) 
(1.1198258018588148, -1.6555400875731) (1.1873794725592695, 
-1.607142857142857) (1.223606685219604, -1.5807495423624605) 
(1.2500000000000004, -1.5588645976457722) (1.3216612386857989, 
-1.5002326672572268) (1.398246348734299, -1.4285714285714284) 
(1.4138825080947321, -1.4138825080947313) (1.428571428571429, 
-1.3982463487342984) (1.5002326672572277, -1.3216612386857984) 
(1.5588645976457727, -1.2499999999999998) (1.5807495423624611, 
-1.2236066852196033) (1.6071428571428577, -1.1873794725592686) 
(1.6555400875731006, -1.119825801858814) (1.6856045180802892, 
-1.0714285714285712) (1.7247626873253183, -1.0104769730396033) 
(1.7857142857142863, -0.9010601002563268) (1.788603096734676, 
-0.8957459538775323) (1.7899411655300907, -0.8928571428571426) 
(1.7939968812887361, -0.8845745472826927) (1.8437382692059525, 
-0.7723096977773803) (1.8684759855853847, -0.714285714285714) 
(1.8750000000000004, -0.69668313788553) (1.8931815687113005, 
-0.6431815687112996) (1.910591247595676, -0.5894087524043242) 
(1.924869599946293, -0.5357142857142854) (1.9341049019638132, 
-0.5055334733923839) (1.9642857142857149, -0.3753810253482636) 
(1.9676135341490355, -0.3604706770061774) (1.9681330321464447, 
-0.3571428571428568) (1.9690470977762486, -0.352381473652323) 
(1.9889801436285635, -0.20326585791427695) (1.9906563620406859, 
-0.17857142857142824) (1.993788841087434, -0.14906830176970898) 
(1.9988845080082551, -0.034598793722540044) (1.9982072874278278, 
  0.001) (1.9988845080082551, 
  0.03459879372254071) (1.9937888410874338, 
  0.14906830176970978) (1.9906563620406856, 
  0.1785714285714289) (1.9889801436285632, 
  0.20326585791427748) (1.9690470977762489, 
  0.3523814736523234) (1.968133032146445, 
  0.3571428571428575) (1.9676135341490355, 
  0.36047067700617824) (1.9642857142857149, 
  0.3753810253482652) (1.9341049019638132, 
  0.5055334733923844) (1.9265899214113738, 
  0.535714285714286) (1.9105912475956754, 
  0.5894087524043253) (1.8931815687113003, 
  0.6431815687113001) (1.8750000000000004, 
  0.6966831378855304) (1.8684759855853847, 
  0.7142857142857146) (1.8437382692059523, 
  0.7723096977773808) (1.7939968812887357, 
  0.8845745472826939) (1.7899411655300905, 
  0.8928571428571432) (1.7886030967346758, 
  0.8957459538775329) (1.7857142857142863, 
  0.901060100256327) (1.724762687325318, 
  1.0104769730396035) (1.6856045180802888, 
  1.0714285714285718) (1.6555400875731003, 
  1.1198258018588145) (1.6071428571428577, 
  1.1873794725592686) (1.5807495423624607, 
  1.2236066852196035) (1.5588645976457722, 
  1.2500000000000004) (1.500232667257227, 
  1.3216612386857987) (1.428571428571429, 
  1.3982463487342984) (1.4138825080947317, 
  1.4138825080947317) (1.3982463487342984, 
  1.428571428571429) (1.3216612386857987, 
  1.500232667257227) (1.2500000000000004, 
  1.5588645976457722) (1.2236066852196035, 
  1.5807495423624607) (1.1873794725592686, 
  1.6071428571428577) (1.1198258018588145, 
  1.6555400875731003) (1.0714285714285718, 
  1.6856045180802888) (1.0104769730396035, 
  1.724762687325318) (0.901060100256327, 
  1.7857142857142863) (0.8957459538775329, 
  1.7886030967346758) (0.8928571428571432, 
  1.7899411655300905) (0.8845745472826939, 
  1.7939968812887357) (0.7723096977773808, 
  1.8437382692059523) (0.7142857142857146, 
  1.8684759855853847) (0.6966831378855304, 
  1.8750000000000004) (0.6431815687113001, 
  1.8931815687113003) (0.5894087524043253, 
  1.9105912475956754) (0.535714285714286, 
  1.9248695999462928) (0.5055334733923844, 
  1.9341049019638132) (0.3753810253482652, 
  1.9642857142857149) (0.3571428571428575, 
  1.968133032146445) (0.3523814736523234, 
  1.9690470977762489) (0.20326585791427748, 
  1.9889801436285632) (0.1785714285714289, 
  1.9906563620406856) (0.14906830176970978, 
  1.9937888410874338) (0.03459879372254071, 
  1.9988845080082551) (0.001, 
  1.999669391322037) (-0.034598793722540044, 
  1.9988845080082551) (-0.14906830176970898, 
  1.993788841087434) (-0.17857142857142824, 
  1.9906563620406859) (-0.20326585791427695, 
  1.9889801436285635) (-0.352381473652323, 
  1.9690470977762486) (-0.3571428571428568, 
  1.9681330321464447) (-0.3604706770061774, 
  1.9676135341490355) (-0.3753810253482636, 
  1.9642857142857149) (-0.5055334733923839, 
  1.9341049019638132) (-0.5357142857142854, 
  1.924869599946293) (-0.5894087524043242, 
  1.910591247595676) (-0.6431815687112996, 
  1.8931815687113005) (-0.714285714285714, 
  1.8648885609350092) (-0.7723096977773803, 
  1.8437382692059525) (-0.8845745472826927, 
  1.7939968812887361) (-0.8928571428571426, 
  1.7899411655300907) (-0.8957459538775323, 
  1.788603096734676) (-0.9010601002563268, 
  1.7857142857142863) (-1.0104769730396033, 
  1.7247626873253183) (-1.0714285714285712, 
  1.6856045180802892) (-1.119825801858814, 
  1.6555400875731006) (-1.1873794725592686, 
  1.6071428571428577) (-1.2236066852196033, 
  1.5807495423624611) (-1.2499999999999998, 
  1.5588645976457727) (-1.3216612386857984, 
  1.5002326672572277) (-1.3982463487342984, 
  1.428571428571429) (-1.4138825080947313, 
  1.4138825080947321) (-1.4285714285714284, 
  1.398246348734299) (-1.5002326672572268, 
  1.3216612386857989) (-1.5588645976457722, 
  1.2500000000000004) (-1.5807495423624605, 
  1.223606685219604) (-1.607142857142857, 
  1.1873794725592695) (-1.6555400875731, 
  1.1198258018588148) (-1.685604518080289, 
  1.0714285714285718) (-1.7247626873253177, 
  1.010476973039604) (-1.7857142857142856, 
  0.9010601002563274) (-1.7886030967346753, 
  0.895745953877533) (-1.78994116553009, 
  0.8928571428571432) (-1.7939968812887355, 
  0.8845745472826934) (-1.843738269205952, 
  0.7723096977773811) (-1.864888560935009, 
  0.7142857142857146) (-1.8931815687113, 
  0.6431815687113004) (-1.9105912475956757, 
  0.5894087524043247) (-1.9248695999462928, 
  0.535714285714286) (-1.934104901963813, 
  0.5055334733923849) (-1.9642857142857142, 
  0.3753810253482671) (-1.9676135341490353, 
  0.3604706770061786) (-1.9681330321464445, 
  0.3571428571428575) (-1.9690470977762489, 
  0.3523814736523229) (-1.9889801436285635, 
  0.20326585791427815) (-1.9906563620406856, 
  0.1785714285714289) (-1.993788841087434, 
  0.14906830176970903) (-1.9988845080082551, 
  0.034598793722541266) (-1.9996693913220367, 
  0.001) (-1.9988845080082551, 
-0.0345987937225406) (-1.9937888410874343, -0.1490683017697082) 
(-1.9906563620406859, -0.17857142857142824) (-1.9889801436285635, 
-0.20326585791427762) (-1.969047097776249, -0.352381473652322) 
(-1.9681330321464447, -0.3571428571428568) (-1.9676135341490355, 
-0.3604706770061781) (-1.9642857142857142, -0.3753810253482673) 
(-1.934104901963813, -0.5055334733923842) (-1.9248695999462928, 
-0.5357142857142854) (-1.9105912475956761, -0.5894087524043236) 
(-1.8931815687113003, -0.6431815687113001) (-1.875, 
-0.6966831378855314) (-1.868475985585385, -0.714285714285714) 
(-1.8437382692059525, -0.7723096977773807) (-1.7939968812887368, 
-0.8845745472826915) (-1.7899411655300907, -0.8928571428571426) 
(-1.7886030967346758, -0.8957459538775328) (-1.7857142857142856, 
-0.9010601002563281) (-1.7247626873253181, -1.0104769730396037) 
(-1.6856045180802892, -1.0714285714285712) (-1.6555400875731001, 
-1.1198258018588145) (-1.607142857142857, -1.1873794725592692) 
(-1.5807495423624607, -1.2236066852196035) (-1.5588645976457727, 
-1.2499999999999998) (-1.5002326672572273, -1.3216612386857984) 
(-1.4285714285714284, -1.398246348734299) (-1.4138825080947317, 
-1.4138825080947317) (-1.398246348734299, -1.4285714285714284) 
(-1.3216612386857984, -1.5002326672572273) (-1.2499999999999998, 
-1.5588645976457727) (-1.2236066852196035, -1.5807495423624607) 
(-1.1873794725592692, -1.607142857142857) (-1.1198258018588145, 
-1.6555400875731001) (-1.0714285714285712, -1.6856045180802892) 
(-1.0104769730396037, -1.7247626873253181) (-0.9010601002563281, 
-1.7857142857142856) (-0.8957459538775328, -1.7886030967346758)};
\end{axis}
\end{tikzpicture}%
}
\end{minipage}\hfill
\begin{minipage}[t]{4.275cm}
     \centering
        \resizebox{\linewidth}{!}{%
\pgfplotsset{every axis/.append style={font=\Large,
extra description/.code={5
}}}
\begin{tikzpicture}
\begin{axis}[axis lines = middle,
    thick,no markers,
   xlabel = {\makebox[0pt]{\hspace{25pt}\Large $\bar{k}_{x}$}},
   ylabel = {\raisebox{15pt}[0pt][0pt]{\Large $\bar{k}_{y}$}},
axis on top]
\addplot[black,smooth,very thick,color=black,fill=blue!10] coordinates{ (-1.58538,-1.79966)(-1.57612,-1.82143)(-1.55324,-1.87467)(-1.55265,-1.87592)(-1.52628,-1.92857)(-1.51746,-1.94603)(-1.5,-1.97748)(-1.46676,-2.03571)(-1.43871,-2.08157)(-1.4073,-2.12841)(-1.39738,-2.14286)(-1.39286,-2.14906)(-1.34829,-2.20543)(-1.31417,-2.25)(-1.28571,-2.28287)(-1.24965,-2.32108)(-1.21653,-2.35714)(-1.17857,-2.39406)(-1.14238,-2.42809)(-1.10029,-2.46429)(-1.07143,-2.48788)(-1.02469,-2.52469)(-0.976241,-2.55947)(-0.964286,-2.56789)(-0.95903,-2.57143)(-0.895811,-2.6101)(-0.857143,-2.63417)(-0.77622,-2.67857)(-0.75901,-2.68758)(-0.75,-2.69181)(-0.724663,-2.70391)(-0.685865,-2.72158)(-0.642857,-2.73889)(-0.609689,-2.75255)(-0.545762,-2.77567)(-0.535714,-2.77904)(-0.514021,-2.78571)(-0.447423,-2.80457)(-0.428571,-2.80997)(-0.39736,-2.81693)(-0.321429,-2.83372)(-0.270351,-2.84178)(-0.214286,-2.85073)(-0.142377,-2.85762)(-0.107143,-2.86097)(-0.0762313,-2.86195)(-0.0001,-2.86438)(0.0762313,-2.86195)(0.107143,-2.86097)(0.142377,-2.85762)(0.214286,-2.85073)(0.270351,-2.84178)(0.321429,-2.83372)(0.39736,-2.81693)(0.428571,-2.80997)(0.447988,-2.80513)(0.514021,-2.78571)(0.530751,-2.78075)(0.535714,-2.77904)(0.545762,-2.77567)(0.609689,-2.75255)(0.642857,-2.73889)(0.685865,-2.72158)(0.724663,-2.70391)(0.75,-2.69181)(0.75901,-2.68758)(0.77622,-2.67857)(0.857143,-2.63417)(0.895811,-2.6101)(0.95903,-2.57143)(0.964286,-2.56789)(1.02337,-2.52337)(1.07143,-2.48788)(1.10029,-2.46429)(1.14133,-2.42705)(1.21329,-2.35714)(1.24965,-2.32108)(1.28571,-2.2783)(1.34829,-2.20543)(1.39324,-2.14286)(1.43732,-2.08018)(1.5,-1.97387)(1.51692,-1.94549)(1.52534,-1.92857)(1.5496,-1.87898)(1.58461,-1.7989)(1.61432,-1.71429)(1.63999,-1.63999)(1.66501,-1.54927)(1.67504,-1.5)(1.68165,-1.46737)(1.70693,-1.29307)(1.70733,-1.28571)(1.70774,-1.27917)(1.71114,-1.07457)(1.71085,-1.07143)(1.71058,-1.06772)(1.69071,-0.880722)(1.68514,-0.857143)(1.67803,-0.820891)(1.65268,-0.704464)(1.63435,-0.642857)(1.62847,-0.621527)(1.60714,-0.560438)(1.6005,-0.542362)(1.58927,-0.517842)(1.55062,-0.428571)(1.53534,-0.39323)(1.5,-0.325571)(1.49748,-0.321429)(1.49396,-0.315387)(1.45895,-0.255337)(1.44643,-0.235079)(1.43232,-0.214286)(1.41683,-0.190316)(1.39598,-0.160714)(1.39286,-0.155982)(1.37218,-0.127823)(1.3556,-0.107143)(1.33929,-0.0849114)(1.32498,-0.0678809)(1.31205,-0.0535714)(1.28571,-0.0211565)(1.27536,-0.0103574)(1.26475,-0.0001)(1.23214,0.0366023)(1.22326,0.0446856)(1.21295,0.0535714)(1.17857,0.089323)(1.16861,0.097177)(1.15553,0.107143)(1.125,0.137557)(1.11117,0.146883)(1.09086,0.160714)(1.07143,0.180955)(1.05008,0.192937)(1.04464,0.197616)(1.02044,0.211704)(1.01786,0.213174)(1.01699,0.213418)(1.01431,0.214286)(0.99877,0.221984)(0.991071,0.226683)(0.977343,0.227343)(0.964286,0.237771)(0.95339,0.225181)(0.9375,0.216746)(0.935298,0.214286)(0.919931,0.205069)(0.910714,0.198898)(0.899305,0.1875)(0.890038,0.181391)(0.866843,0.160714)(0.857143,0.155775)(0.829981,0.134305)(0.803571,0.114166)(0.80025,0.110464)(0.796306,0.107143)(0.776786,0.093883)(0.769161,0.0879822)(0.75,0.0767918)(0.7374,0.0661711)(0.719676,0.0535714)(0.704905,0.0450951)(0.696429,0.0410502)(0.671613,0.0248156)(0.662983,0.0201261)(0.642857,0.00923891)(0.63747,0.00538679)(0.628343,-0.0001)(0.602506,-0.0132207)(0.589286,-0.0187831)(0.566645,-0.0309307)(0.551903,-0.0373827)(0.535714,-0.0445265)(0.51678,-0.0535714)(0.492308,-0.0637364)(0.482143,-0.0669707)(0.461873,-0.0738417)(0.428571,-0.0866591)(0.413945,-0.0925164)(0.375,-0.104322)(0.365686,-0.107143)(0.331188,-0.116902)(0.321429,-0.118776)(0.306736,-0.121835)(0.243714,-0.136571)(0.214286,-0.141308)(0.174742,-0.146687)(0.150454,-0.150454)(0.107143,-0.154617)(0.101503,-0.155074)(0.05656,-0.157726)(0.0535714,-0.157901)(0.0508324,-0.157975)(-0.0001,-0.15899)(-0.0508324,-0.157975)(-0.0566263,-0.157659)(-0.107143,-0.154617)(-0.150454,-0.150454)(-0.174742,-0.146687)(-0.214286,-0.141308)(-0.243714,-0.136571)(-0.306736,-0.121835)(-0.321429,-0.118929)(-0.331188,-0.116902)(-0.365686,-0.107143)(-0.375,-0.104322)(-0.379129,-0.103014)(-0.413789,-0.0923605)(-0.428571,-0.0866591)(-0.461873,-0.0738417)(-0.492308,-0.0637364)(-0.51678,-0.0535714)(-0.535714,-0.0445265)(-0.551903,-0.0373827)(-0.566645,-0.0309307)(-0.575411,-0.0267857)(-0.589286,-0.0196324)(-0.602506,-0.0132207)(-0.616071,-0.00591514)(-0.627831,-0.0001)(-0.63747,0.00538679)(-0.642857,0.00923891)(-0.662983,0.0201261)(-0.671613,0.0248156)(-0.696429,0.0410502)(-0.704905,0.0450951)(-0.719676,0.0535714)(-0.7374,0.0661711)(-0.75,0.0767918)(-0.769161,0.0879822)(-0.776786,0.093883)(-0.796306,0.107143)(-0.80025,0.110464)(-0.803571,0.114166)(-0.827995,0.131567)(-0.830732,0.133554)(-0.857143,0.155775)(-0.866843,0.160714)(-0.890038,0.181391)(-0.899305,0.1875)(-0.910714,0.198898)(-0.919931,0.205069)(-0.935298,0.214286)(-0.9375,0.216746)(-0.95339,0.225181)(-0.964286,0.229282)(-0.977343,0.227343)(-0.991071,0.226683)(-1.01431,0.214286)(-1.01699,0.213418)(-1.01786,0.213174)(-1.02044,0.211704)(-1.05008,0.192937)(-1.0568,0.1875)(-1.07143,0.178513)(-1.0812,0.170486)(-1.09334,0.160714)(-1.09821,0.157529)(-1.11117,0.146883)(-1.125,0.137557)(-1.15553,0.107143)(-1.16868,0.0972559)(-1.17857,0.0911023)(-1.22332,0.0447453)(-1.2627,-0.0001)(-1.27541,-0.0103028)(-1.28571,-0.0199481)(-1.32503,-0.0678264)(-1.35322,-0.107143)(-1.37218,-0.127823)(-1.39286,-0.15436)(-1.43048,-0.214286)(-1.45895,-0.255337)(-1.46606,-0.267857)(-1.49396,-0.315387)(-1.49748,-0.321429)(-1.5,-0.325571)(-1.53534,-0.39323)(-1.55062,-0.428571)(-1.5694,-0.466318)(-1.57591,-0.482143)(-1.5911,-0.519668)(-1.59784,-0.535714)(-1.60054,-0.542314)(-1.60714,-0.560438)(-1.63435,-0.642857)(-1.65268,-0.704464)(-1.66377,-0.75)(-1.67803,-0.820891)(-1.68646,-0.857143)(-1.69071,-0.880722)(-1.70222,-0.964286)(-1.71058,-1.06772)(-1.71099,-1.07143)(-1.71114,-1.07457)(-1.71278,-1.17857)(-1.70774,-1.27917)(-1.7076,-1.28571)(-1.70693,-1.29307)(-1.69549,-1.39286)(-1.68165,-1.46737)(-1.67648,-1.5)(-1.66501,-1.54927)(-1.6506,-1.60714)(-1.63999,-1.63999)(-1.61787,-1.71429)(-1.61511,-1.72225)(-1.60714,-1.74295)(-1.58538,-1.79966)};
\addplot[black,smooth,very thick,color=black,fill=blue!10] coordinates{(1.03407577826829, 0.35878136458885224) (1.0178571428571426, 
  0.3373664914455965) (0.9887315421365942, 
  0.32142857142857106) (0.9782526942915686, 
  0.30746159142271645) (0.964285714285714, 
  0.29328378892610124) (0.9460725053376466, 
  0.3032153624805037) (0.9264012375973719, 
  0.3057416195454845) (0.9107142857142854, 
  0.30882888689431753) (0.8980472809658249, 
  0.32142857142857106) (0.869246014710431, 
  0.3335317289961452) (0.8571428571428568, 
  0.3365872380441026) (0.8336831248647953, 
  0.3515402677219383) (0.8218212527870952, 
  0.35675017578433266) (0.8035714285714282, 
  0.36521343515928484) (0.7981264206713689, 
  0.36955499209994036) (0.7888187721399883, 
  0.37499999999999967) (0.7499999999999996, 
  0.39129443810382064) (0.7261968522140881, 
  0.4047682807855168) (0.696428571428571, 
  0.41593246433350856) (0.6703003740927143, 
  0.4285714285714282) (0.650812647937237, 
  0.4365269336515228) (0.6428571428571425, 
  0.4381184432816302) (0.6279646320039826, 
  0.44346393942458817) (0.5731559899139322, 
  0.4660131327710751) (0.5357142857142854, 
  0.4758200274917203) (0.4919822506278368, 
  0.4919822506278368) (0.4666601476358393, 
  0.4976255666498743) (0.4285714285714282, 
  0.5080165563355391) (0.40777076631777964, 
  0.5149136234606367) (0.37499999999999967, 
  0.5216119315019863) (0.32544345874279573, 
  0.5316993984000606) (0.32142857142857106, 
  0.5327310201626142) (0.31903124185594667, 
  0.533316956141661) (0.30405651752713725, 
  0.5357142857142854) (0.22696326806657383, 
  0.5483918394951453) (0.21428571428571394, 
  0.5499351060166054) (0.1989428535332219, 
  0.5510571464667774) (0.12986715793854764, 
  0.5584385865099762) (0.1071428571428568, 
  0.5603537665443413) (0.0820206424702142, 
  0.560836500386928) (0.027153472262797733, 
  0.5628677579770834) (-0.0001, 
  0.5638305961813682) (-0.0820599411640737, 
  0.5607972016930691) (-0.13063966403243218, 
  0.55921109260386) (-0.2142857142857146, 
  0.5499351060166053) (-0.22696326806657444, 
  0.5483918394951453) (-0.3040565175271371, 
  0.5357142857142854) (-0.3190312418559472, 
  0.5333169561416609) (-0.32142857142857173, 
  0.5326941763520652) (-0.3254434587427966, 
  0.5316993984000604) (-0.40777076631778014, 
  0.5149136234606366) (-0.4285714285714289, 
  0.508016556335539) (-0.4666601476358403, 
  0.49762556664987395) (-0.49198225062783724, 
  0.49198225062783657) (-0.535714285714286, 
  0.4758200274917199) (-0.5731559899139326, 
  0.46601313277107476) (-0.6279646320039836, 
  0.4434639394245878) (-0.6428571428571431, 
  0.43811844328163) (-0.6508126479372376, 
  0.4365269336515226) (-0.6737738248748005, 
  0.4285714285714282) (-0.7261968522140887, 
  0.4047682807855167) (-0.7500000000000002, 
  0.38582817800485464) (-0.798251319821881, 
  0.36967989125045186) (-0.8268440947112509, 
  0.3517273338601777) (-0.8571428571428574, 
  0.3365872380441026) (-0.8697928999336808, 
  0.33407861421939444) (-0.8980472809658255, 
  0.32142857142857106) (-0.910714285714286, 
  0.30882888689431726) (-0.9264012375973726, 
  0.3057416195454845) (-0.9460725053376478, 
  0.30321536248050435) (-0.9642857142857145, 
  0.30191580671773444) (-0.9782526942915691, 
  0.3074615914227164) (-0.9887315421365932, 
  0.32142857142857106) (-1.017857142857143, 
  0.3373664914455965) (-1.0340757782682903, 
  0.35878136458885224) (-1.0440510037588264, 
  0.37499999999999967) (-1.0714285714285716, 
  0.4041182122739693) (-1.0810983740160272, 
  0.41890162598397257) (-1.0839056332823955, 
  0.4285714285714282) (-1.1135091666354693, 
  0.47065202377832577) (-1.1231696538733196, 
  0.48397320326953736) (-1.1250000000000002, 
  0.48677838610822677) (-1.1517692306772718, 
  0.5357142857142854) (-1.1621553277270316, 
  0.5521303865586827) (-1.1785714285714288, 
  0.5828511280551704) (-1.2025332978860153, 
  0.6428571428571425) (-1.2210271818522038, 
  0.6853128961379173) (-1.2278618095997187, 
  0.7007096189717098) (-1.2425935231927872, 
  0.7499999999999996) (-1.2546985349927486, 
  0.7810157507215368) (-1.2700233614335728, 
  0.8414519328621436) (-1.2740421982510384, 
  0.8571428571428568) (-1.276805039726765, 
  0.8660521031303778) (-1.285714285714286, 
  0.9119871380009559) (-1.2945612794924946, 
  0.964285714285714) (-1.3045390050579069, 
  1.0526038520849503) (-1.3052965336661815, 
  1.0714285714285712) (-1.3047751860659307, 
  1.0904894717802158) (-1.303070643740422, 
  1.2683579276881494) (-1.3013579560839112, 
  1.2857142857142854) (-1.2989228461418272, 
  1.2989228461418265) (-1.2865910414236406, 
  1.3928571428571423) (-1.285714285714286, 
  1.396739488300339) (-1.2650328813844227, 
  1.4793185956701362) (-1.2609722692048204, 
  1.4999999999999996) (-1.2499460109836682, 
  1.5357682747306174) (-1.2114578658125805, 
  1.6400292943840085) (-1.1767667232595502, 
  1.714285714285714) (-1.1423396230660379, 
  1.7851967659231802) (-1.0714285714285716, 
  1.8994689156330458) (-1.0604738785878631, 
  1.9176167357307197) (-1.052068717768344, 
  1.9285714285714282) (-0.9728658111380827, 
  2.027134188861917) (-0.9650766473431291, 
  2.0365052187717) (-0.9046657989658372, 
  2.0953342010341625) (-0.8571428571428574, 
  2.141390466201518) (-0.8564144352560276, 
  2.1421287209703124) (-0.8554859658505859, 
  2.1428571428571423) (-0.7341886135274581, 
  2.2341886135274573) (-0.6428571428571431, 
  2.286335859415985) (-0.5988301074566565, 
  2.31311582174237) (-0.535714285714286, 
  2.3459947981912697) (-0.5095050226009088, 
  2.3571428571428568) (-0.45068703800405363, 
  2.3792584665754815) (-0.4285714285714289, 
  2.3877908868110684) (-0.3815917346651085, 
  2.404122551049177) (-0.3715333345229712, 
  2.407247620237256) (-0.32142857142857173, 
  2.4196459561235124) (-0.2861148973813943, 
  2.4289720402385364) (-0.2401861271457084, 
  2.4383853014257197) (-0.2142857142857146, 
  2.442594890509682) (-0.19589675121246214, 
  2.445896751212461) (-0.11563899262330199, 
  2.455789578805269) (-0.10714285714285747, 
  2.456444226594611) (-0.09997718209968336, 
  2.4571200392425396) (-0.0031839212214766835, 
  2.461101793064237) (-0.0001, 
  2.461074190930411) (0.0031839212214760174, 
  2.461101793064237) (0.09997718209968269, 
  2.4571200392425396) (0.1071428571428568, 
  2.456444226594611) (0.11563899262330124, 
  2.4557895788052693) (0.19589675121246164, 
  2.4458967512124614) (0.21428571428571394, 
  2.442594890509682) (0.24018612714570772, 
  2.4383853014257197) (0.2861148973813938, 
  2.4289720402385364) (0.32142857142857106, 
  2.419645956123513) (0.37153333452297055, 
  2.407247620237256) (0.3815917346651078, 
  2.404122551049177) (0.4285714285714282, 
  2.3877908868110684) (0.450687038004053, 
  2.3792584665754815) (0.5095050226009091, 
  2.3571428571428568) (0.5357142857142854, 
  2.3459947981912705) (0.5563300437630353, 
  2.336527099094107) (0.6005461340151149, 
  2.314831848300829) (0.6428571428571425, 
  2.2908820173813575) (0.6692518043612886, 
  2.2763946615041455) (0.7111631314921537, 
  2.2499999999999996) (0.7348264869910897, 
  2.2348264869910897) (0.7499999999999996, 
  2.223733604723522) (0.7972346823441547, 
  2.1900918252012977) (0.8555991632501646, 
  2.1428571428571423) (0.8564448665693707, 
  2.1421591522836563) (0.8571428571428568, 
  2.141507399730792) (0.9046657989658256, 
  2.0953342010341736) (0.964285714285714, 
  2.0373686454179105) (0.9650766473431289, 
  2.0365052187717003) (0.9658365575989999, 
  2.035714285714285) (0.976155506089027, 
  2.023844493910972) (1.0146854450522975, 
  1.9789711593380117) (1.0178571428571426, 
  1.9750742129978014) (1.0536426873287457, 
  1.9285714285714282) (1.060962841285611, 
  1.918105698428468) (1.0714285714285712, 
  1.9023797988078428) (1.1040474981894188, 
  1.8540474981894186) (1.1236442581972796, 
  1.8214285714285712) (1.1440017692161006, 
  1.7868589120732434) (1.1785714285714284, 
  1.721061358866901) (1.1809200614928426, 
  1.7166343472071282) (1.1820293771400032, 
  1.714285714285714) (1.1850334506224032, 
  1.707823692234739) (1.2131563196728952, 
  1.6417277482443235) (1.2272928101372151, 
  1.6071428571428568) (1.2321428571428568, 
  1.5933429026520405) (1.242107247765888, 
  1.5635358191944595) (1.2516463327999527, 
  1.5340679529143322) (1.2609722692048204, 
  1.4999999999999996) (1.2662369917513117, 
  1.4805227060370258) (1.2857142857142854, 
  1.3967394883003417) (1.2864465915571688, 
  1.3935894487000258) (1.2866135000188188, 
  1.3928571428571423) (1.2868113482131007, 
  1.391760080358327) (1.2997878768922135, 
  1.2997878768922135) (1.3017787926528903, 
  1.2857142857142854) (1.303493094585627, 
  1.2679354768429438) (1.307000661898923, 
  1.1998578047560657) (1.307592328814997, 
  1.1785714285714284) (1.3085806478264046, 
  1.1557050664593091) (1.3063807531935092, 
  1.092095038907795) (1.3052965336661815, 
  1.0714285714285712) (1.3047726046094765, 
  1.05237025253338) (1.295843851155023, 
  0.9744152797264517) (1.2945612794924946, 
  0.964285714285714) (1.2938421376565699, 
  0.9561578623434296) (1.2857142857142854, 
  0.9119871380009515) (1.2768446716318567, 
  0.8660124712252854) (1.2740421982510386, 
  0.8571428571428568) (1.2700233614335732, 
  0.8414519328621446) (1.2546985349927484, 
  0.7810157507215365) (1.2445264240415896, 
  0.7499999999999996) (1.2321428571428568, 
  0.712508591068223) (1.2278618095997182, 
  0.7007096189717095) (1.221027181852204, 
  0.6853128961379182) (1.202533297886015, 
  0.6428571428571425) (1.1785714285714284, 
  0.5828511280551697) (1.1621553277270313, 
  0.5521303865586824) (1.1517692306772715, 
  0.5357142857142854) (1.1249999999999998, 
  0.48677838610822677) (1.1231696538733191, 
  0.48397320326953736) (1.1135091666354695, 
  0.4706520237783266) (1.0855520223966395, 
  0.4285714285714282) (1.0807821998622384, 
  0.419217800137761) (1.0714285714285712, 
  0.40411821227396816) (1.044051003758826, 
  0.37499999999999967) (1.03407577826829, 0.35878136458885224)};
\end{axis}
\end{tikzpicture}%
}
\end{minipage}
\caption{Fermi surface topology for two dimensional electronic systems under Rashba interaction effect in the low density regime, with $\Delta_{0}/2E_{0}=0.1$ and $J=1$. (\emph{Left}) $\bar{E}_{x}=0$, (\emph{Right}) $\bar{E}_{x}=1.15$. Calculations in both cases are performed under the constraint $N_{-}/\mathcal{S}=2k_{0}^{2}/\pi$=constant.}\label{AAA}
\end{figure}
\section{Results and Discussion}
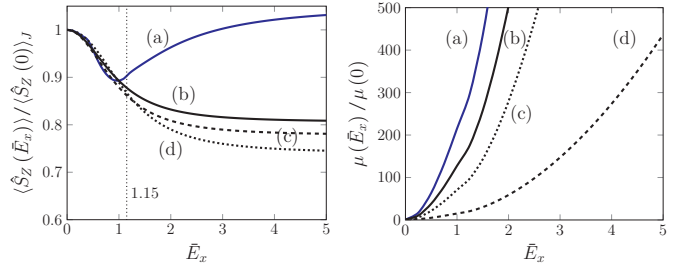
\begin{figure}[!h]
\begin{minipage}[b]{0.5\columnwidth}
     \centering
        \resizebox{\linewidth}{!}{%
\centering
\pgfplotsset{every axis/.append style={font=\large,
extra description/.code={5
\node at (0.4,0.325) {\Large{(d)}};
\node at (0.3,0.125){1.15};
\node at (0.45,0.585) {\Large{(b)}};
\node at (0.85,0.385) {\Large{(c)}};
\node at (0.35,0.85) {\Large{(a)}};
}}}
\begin{tikzpicture}[scale=1,line width=3pt]
\begin{axis}[
scale=1,
domain=0.0:5,xmax=5,xmin=0,ymax=1.05, ymin=0.6,
xlabel={\Large{$\bar{E}_{x}$}},
ylabel={\Large{$\langle \hat{S}_{Z}\left(\bar{E}_{x}\right)\rangle/\langle\hat{S}_{Z}\left(0\right)\rangle_{J}$}}
]
\addplot[blue,ultra thick,smooth] coordinates {(0., 1.) (0.1, 0.9983538116294702) (0.2, 
  0.9936688868445283) (0.30000000000000004, 0.9866742407911635) (0.4, 
  0.9783599574932373) (0.5, 0.9573296000676084) (0.6000000000000001, 
  0.9337602242821215) (0.7000000000000001, 0.914754736107196) (0.8, 
  0.9014370123048908) (0.9, 0.8942127417119483) (1., 
  0.8930338321726037) (1.1, 0.8978234680003726) (1.2000000000000002, 
  0.9084838565808343) (1.3, 0.9167250915668904) (1.4000000000000001, 
  0.924117092582441) (1.5, 0.9312383786964211) (1.6, 
  0.9381638856419194) (1.7000000000000002, 0.9448561300728882) (1.8, 
  0.9512622740233311) (1.9000000000000001, 0.9573410418499022) (2., 
  0.9630680505598438) (2.1, 0.9684340507816522) (2.2, 
  0.9734413454358385) (2.3000000000000003, 
  0.9781004003885803) (2.4000000000000004, 0.9824270264037217) (2.5, 
  0.9864398685374264) (2.6, 0.9901591877539886) (2.7, 
  0.9936056556059973) (2.8000000000000003, 
  0.9967996439117609) (2.9000000000000004, 0.9997607426869574) (3., 
  1.0025075096450975) (3.1, 1.0050573205719437) (3.2, 
  1.0074262722927823) (3.3000000000000003, 
  1.0096292452966469) (3.4000000000000004, 1.0116798886654905) (3.5, 
  1.013590725077188) (3.6, 1.0153731468930873) (3.7, 
  1.0170375894636308) (3.8000000000000003, 
  1.018593551331071) (3.9000000000000004, 1.020049677535634) (4., 
  1.021413859841178) (4.1000000000000005, 1.0226932647110558) (4.2, 
  1.0238944339755771) (4.3, 1.0250233280406102) (4.4, 
  1.0260853798499918) (4.5, 1.0270855548771949) (4.6000000000000005, 
  1.028028385885968) (4.7, 1.0289180185553235) (4.800000000000001, 
  1.029758246478532) (4.9, 1.0305525447460344) (5., 
  1.0313040998108383)};
\addplot[black,ultra thick,smooth] coordinates{(0., 1.) (0.1, 0.9979320398943122) (0.2, 
  0.9919447914476658) (0.30000000000000004, 0.9826383549281128) (0.4, 
  0.9708633140027167) (0.5, 0.9575555684652451) (0.6000000000000001, 
  0.9435905216704085) (0.7000000000000001, 0.9296869520228236) (0.8, 
  0.9163689691104026) (0.9, 0.9039719814306654) (1., 
  0.8926786354422565) (1.1, 0.8825375452043864) (1.2000000000000002, 
  0.8735428123516823) (1.3, 0.8656232184644189) (1.4000000000000001, 
  0.8586873844283209) (1.5, 0.8526318940129389) (1.6, 
  0.8473541820421334) (1.7000000000000002, 0.8427569292899675) (1.8, 
  0.8387510553142403) (1.9000000000000001, 0.8352570784726946) (2., 
  0.8322050649085163) (2.1, 0.8295343053577009) (2.2, 
  0.8271920163700046) (2.3000000000000003, 
  0.8251334865917468) (2.4000000000000004, 0.823319830167674) (2.5, 
  0.8217180216691738) (2.6, 0.8202993845123784) (2.7, 
  0.8190409226934311) (2.8000000000000003, 
  0.8179207348147033) (2.9000000000000004, 0.8169262817277185) (3., 
  0.8160279972188531) (3.1, 0.8152271131457767) (3.2, 
  0.814507613216078) (3.3000000000000003, 
  0.8138597381080573) (3.4000000000000004, 0.813275070845639) (3.5, 
  0.8127462938750064) (3.6, 0.8122670950603624) (3.7, 
  0.8118319526676512) (3.8000000000000003, 
  0.8114360027057754) (3.9000000000000004, 0.8110750689772332) (4., 
  0.8107454111567556) (4.1000000000000005, 0.8104438267874875) (4.2, 
  0.8101674484977152) (4.3, 0.8099137095009933) (4.4, 
  0.8096804190578727) (4.5, 0.8094655559159745) (4.6000000000000005, 
  0.8092674058515235) (4.7, 0.8090843792914627) (4.800000000000001, 
  0.8089149928569926) (4.9, 0.808758211138483) (5., 
  0.8086127930335545)};
\addplot[black,ultra thick,smooth,dashed] coordinates {(0., 1.) (0.1, 0.9977270256543922) (0.2, 
  0.9911295339121563) (0.30000000000000004, 0.9808265384193731) (0.4, 
  0.9677115174206371) (0.5, 0.9527907701609903) (0.6000000000000001, 
  0.9370300251268029) (0.7000000000000001, 0.9212440956428175) (0.8, 
  0.9060439448162172) (0.9, 0.8918331500057822) (1., 
  0.8788363230432027) (1.1, 0.8671408777452064) (1.2000000000000002, 
  0.8567391498397657) (1.3, 0.8475639285011077) (1.4000000000000001, 
  0.8395153527657689) (1.5, 0.8324797611549711) (1.6, 
  0.8263411751953307) (1.7000000000000002, 0.820989002742248) (1.8, 
  0.8163215722397871) (1.9000000000000001, 0.8122475459759946) (2., 
  0.8086865012788138) (2.1, 0.8055682743472469) (2.2, 
  0.8028322534279725) (2.3000000000000003, 
  0.8004261986099762) (2.4000000000000004, 0.7983053780272755) (2.5, 
  0.7964313011723244) (2.6, 0.7947712761963478) (2.7, 
  0.7932971665778833) (2.8000000000000003, 
  0.7919848911281248) (2.9000000000000004, 0.7908138600668672) (3., 
  0.7897662235626881) (3.1, 0.7888267073409048) (3.2, 
  0.7879829116907872) (3.3000000000000003, 
  0.7872224950225329) (3.4000000000000004, 0.7865361010475593) (3.5, 
  0.7859151505615921) (3.6, 0.785352192014516) (3.7, 
  0.7848408446901646) (3.8000000000000003, 
  0.7843754579797557) (3.9000000000000004, 0.7839510776136831) (4., 
  0.7835634276229693) (4.1000000000000005, 0.7832086539559038) (4.2, 
  0.7828834015683482) (4.3, 0.7825848249594539) (4.4, 
  0.7823101907459827) (4.5, 0.782057223169111) (4.6000000000000005, 
  0.7818238553607246) (4.7, 0.7816082265087744) (4.800000000000001, 
  0.7814087752336302) (4.9, 0.7812239908409471) (5., 
  0.7810525862492642)};
\addplot[black,ultra thick,smooth,dotted] coordinates {(0., 0.9999999999999999) (0.1, 0.9986159134939591) (0.2, 
  0.994462577390578) (0.30000000000000004, 0.9875626686932811) (0.4, 
  0.9780228184088972) (0.5, 0.9661017898446417) (0.6000000000000001, 
  0.9522266831331357) (0.7000000000000001, 0.9369743757647119) (0.8, 
  0.9209891535109379) (0.9, 0.9048935545930559) (1., 
  0.889214318262567) (1.1, 0.8743430624092503) (1.2000000000000002, 
  0.860531371101032) (1.3, 0.8479086081990126) (1.4000000000000001, 
  0.8365109114368064) (1.5, 0.8263075855551819) (1.6, 
  0.8172307812041114) (1.7000000000000002, 0.8091869138986411) (1.8, 
  0.802076192303389) (1.9000000000000001, 0.7957979335922161) (2., 
  0.7902563002860606) (2.1, 0.7853625766738079) (2.2, 
  0.7810370230676792) (2.3000000000000003, 
  0.7772085191903625) (2.4000000000000004, 0.7738145070522353) (2.5, 
  0.7708000698799514) (2.6, 0.76811760399208) (2.7, 
  0.7657255999038524) (2.8000000000000003, 
  0.763588202117434) (2.9000000000000004, 0.7616741720454232) (3., 
  0.7599564697623468) (3.1, 0.7584118048251071) (3.2, 
  0.7570197007476316) (3.3000000000000003, 
  0.755762394688816) (3.4000000000000004, 0.7546218214307757) (3.5, 
  0.7535935503866829) (3.6, 0.7526565690901301) (3.7, 
  0.7518036566399858) (3.8000000000000003, 
  0.7510260015824219) (3.9000000000000004, 0.7503156392353653) (4., 
  0.7496654752833383) (4.1000000000000005, 0.7490698343374464) (4.2, 
  0.7485228496623818) (4.3, 0.7480198443657001) (4.4, 
  0.7475566025492724) (4.5, 0.7471293370263302) (4.6000000000000005, 
  0.7467346784727876) (4.7, 0.7463696397639499) (4.800000000000001, 
  0.7460315212946849) (4.9, 0.7457179239714506) (5., 
  0.7454268328837779)};
\draw[black,thick,dotted] (axis cs:1.15,0.0) -- (axis cs:1.15,1.05);
\end{axis}
\end{tikzpicture}
}
\end{minipage}\hfill
\begin{minipage}[b]{0.5\columnwidth}
     \centering
        \resizebox{\linewidth}{!}{%
\centering
\pgfplotsset{every axis/.append style={font=\large,
extra description/.code={5
\node at (0.85,0.85) {\Large{(d)}};
\node at (0.45,0.5) {\Large{(c)}};
\node at (0.2,0.85) {\Large{(a)}};
\node at (0.425,0.85) {\Large{(b)}};
}}}
\begin{tikzpicture}[scale=1,line width=3pt]
\begin{axis}[
scale=1,
domain=0:5,xmax=5, xmin=0,ymax=500, ymin=0.0,
xlabel={\Large{$\bar{E}_{x}$}},
ylabel={\Large{$\mu\left(\bar{E}_{x}\right)/\mu\left(0\right)$}}
]
\addplot[blue,ultra thick,smooth] coordinates {(0., 1.) (0.25, 17.225337353623253) (0.5, 63.64877274763046) (0.75, 
  132.14850966714332) (1., 215.02867101038106) (1.25, 
  297.9630256463731) (1.5, 435.4263119974207) (1.75, 
  622.4663937182731) (2., 851.2687416641329) (2.25, 
  1118.290752029068) (2.5, 1421.611116188723) (2.75, 
  1760.1022378618363) (3., 2133.0656851877525) (3.25, 
  2540.050230011939) (3.5, 2980.754042491751) (3.75, 
  3454.969191455005) (4., 3962.5487497656695) (4.25, 
  4503.386571636776) (4.5, 5077.404462804535) (4.75, 
  5684.543813847898) (5., 6324.760009692756)};
\addplot[black,ultra thick,smooth] coordinates {(0., 1.) (0.25, 10.571219532542433) (0.5, 38.22633418006156) (0.75, 
  78.57344556211476) (1., 127.47136949348898) (1.25, 
  176.4804692977404) (1.5, 257.8403609571007) (1.75, 
  368.56156617403224) (2., 504.01577038804083) (2.25, 
  662.1022602415134) (2.5, 841.6816787886332) (2.75, 
  1042.0854582526217) (3., 1262.899566045311) (3.25, 
  1503.8565579560561) (3.5, 1764.777564341781) (3.75, 
  2045.5393763783309) (4., 2346.0549416836516) (4.25, 
  2666.261375567049) (4.5, 3006.1123561288628) (4.75, 
  3365.5731655152367) (5., 3744.6173769128072)};
\addplot[black,ultra thick,smooth,dashed] coordinates {(0., 1.) (0.25, 2.0438293521674242) (0.5, 5.113299558921042) (0.75, 
  9.87284610944856) (1., 15.192371937505236) (1.25, 
  20.712306020112216) (1.5, 30.143201415912877) (1.75, 
  43.01756319619679) (2., 58.79053543253008) (2.25, 
  77.21017655824629) (2.5, 98.1399047572029) (2.75, 
  121.49985303943889) (3., 147.24073821882746) (3.25, 
  175.33082410093587) (3.5, 205.74893918352234) (3.75, 
  238.4805302549546) (4., 273.5153322911533) (4.25, 
  310.84594037669956) (4.5, 350.4669062287805) (4.75, 
  392.3741503583892) (5., 436.5645700454769)};
\addplot[black,ultra thick,smooth,dotted] coordinates {(0., 0.9999999999999999) (0.25, 6.264691268062313) (0.5, 
  21.634209023605532) (0.75, 43.9013727568326) (1., 
  70.81731510447196) (1.25, 97.88236701646369) (1.5, 
  142.9472799310906) (1.75, 204.29488409676313) (2., 
  279.35796116977906) (2.25, 366.9687564465843) (2.5, 
  466.4938999767115) (2.75, 577.5618573958949) (3., 
  699.9426273373098) (3.25, 833.4876883302619) (3.5, 
  978.0977344145317) (3.75, 1133.7043770679304) (4., 
  1300.259305789186) (4.25, 1477.7276275368595) (4.5, 
  1666.083642846274) (4.75, 1865.3080922994454) (5., 
  2075.3863172206065)};
\end{axis}
\end{tikzpicture}
}
\end{minipage}
\caption{(\emph{Left.}) Average $Z$-spin density as a function of $\bar{E}_{x}$ at zero temperature and low carrier density at $\bar{\Delta}_{0}=0.1$, with $\langle \hat{S}_{Z}(0)\rangle_{J}=-2m^{\star}\Delta_{0}/\pi\hbar^{2}$.  (\emph{Right.}) Normalized Fermi energy behavior as a function of $\bar{E}_{x}$.  In all cases, (a) $J=0$, (b) $J=0.25$, (c) $J=0.5$ and (d) $J=1$. The total number of particles on the low energy sub band $N_{-}$ is conserved.}\label{muu}
\end{figure}

\begin{figure}[h]
\begin{minipage}[b]{0.5\columnwidth}
     \centering
        \resizebox{\linewidth}{!}{%
\centering
\pgfplotsset{every axis/.append style={font=\large,
extra description/.code={5
\node at (0.125,0.935) {\bf{(a)}};
\node at (0.125,0.8) {\bf{(b)}};
\node at (0.25,0.75) {\bf{(c)}};
\node at (0.125,0.525){\bf{(d)}};
\node at (0.5,0.125){1.15};
}}}
\begin{tikzpicture}[scale=1,line width=3pt]
\begin{axis}[
scale=1,
domain=0:2,xmax=2, xmin=0,ymax=1.9, ymin=1,
xlabel={\Large{$\bar{E}_{x}$}},
ylabel={\Large{$\mathcal{D}_{ZZ}^{0}\left(\bar{E}_{x}\right)$ (a.u.)}}
]
\addplot[blue,ultra thick,smooth] coordinates {(0., 1.8094848952758569) (0.1, 1.8099879054792125) (0.2, 
  1.8094925174847367) (0.30000000000000004, 1.8092853832354874) (0.4, 
  1.8089184962006752) (0.5, 1.8372178364391099) (0.6000000000000001, 
  1.806314245401849) (0.7000000000000001, 1.7629404583364208) (0.8, 
  1.7059279313085098) (0.9, 1.6433235735002836) (1., 
  1.5848913968093667) (1.1, 1.5393383587253322) (1.2000000000000002, 
  1.422584825161108) (1.3, 1.3351468995869116) (1.4000000000000001, 
  1.2607777147672872) (1.5, 1.1950567244598802) (1.6, 
  1.1335437105845345) (1.7000000000000002, 1.0774520835301675) (1.8, 
  1.0254892203828547) (1.9000000000000001, 0.9782593198119501) (2., 
  0.9346976798910516) (2.1, 0.8947485502177424) (2.2, 
  0.8584552259966394) (2.3000000000000003, 
  0.8242420017524879) (2.4000000000000004, 0.7927094605618018) (2.5, 
  0.7633894447533497) (2.6, 0.7363785910547145) (2.7, 
  0.7110323137628429) (2.8000000000000003, 
  0.6873506229941979) (2.9000000000000004, 0.6656300367801281) (3., 
  0.6450504356411795) (3.1, 0.6257949459923124) (3.2, 
  0.605393835100578) (3.3000000000000003, 
  0.5883528738461888) (3.4000000000000004, 0.5718374512045953) (3.5, 
  0.5567330341422945) (3.6, 0.5398600538319641) (3.7, 
  0.5235392635942488) (3.8000000000000003, 
  0.5107132763251775) (3.9000000000000004, 0.5004298802616279) (4., 
  0.4884327779214385) (4.1000000000000005, 0.47760899378966354) (4.2, 
  0.4673680379439152) (4.3, 0.4559465160187261) (4.4, 
  0.4467076477864959) (4.5, 0.43358196899577034) (4.6000000000000005, 
  0.4250316423702564) (4.7, 0.4169076527008431) (4.800000000000001, 
  0.4070612080708311) (4.9, 0.3946578528786676) (5., 
  0.38744000287336794)};
\addplot[black,ultra thick,smooth] coordinates {(0., 1.7679581047370851) (0.1, 1.7687238424024057) (0.2, 
  1.7685392542640508) (0.30000000000000004, 1.76789367341839) (0.4, 
  1.7672584210706626) (0.5, 1.8122140151995614) (0.6000000000000001, 
  1.7898987120127599) (0.7000000000000001, 1.7524402153268097) (0.8, 
  1.699416692232437) (0.9, 1.634271798769867) (1., 
  1.5829516648573503) (1.1, 1.5390865184442732) (1.2000000000000002, 
  1.4238886776755229) (1.3, 1.3330034079585618) (1.4000000000000001, 
  1.258468631827574) (1.5, 1.19220392307363) (1.6, 
  1.1304853962717254) (1.7000000000000002, 1.0741895597821525) (1.8, 
  1.0226157022185913) (1.9000000000000001, 0.9750050158898032) (2., 
  0.9319513651912376) (2.1, 0.8925572908585054) (2.2, 
  0.8557122934582025) (2.3000000000000003, 
  0.8215455981330657) (2.4000000000000004, 0.7901406043864823) (2.5, 
  0.7551372074038237) (2.6, 0.7342230816368556) (2.7, 
  0.7079241229765735) (2.8000000000000003, 
  0.684539001299115) (2.9000000000000004, 0.6636449804785923) (3., 
  0.6434153067505666) (3.1, 0.6210990842074688) (3.2, 
  0.603294573801654) (3.3000000000000003, 
  0.5866305121097612) (3.4000000000000004, 0.572540939464888) (3.5, 
  0.5541680160322267) (3.6, 0.5371702757417058) (3.7, 
  0.5240511800142323) (3.8000000000000003, 
  0.5117125612299243) (3.9000000000000004, 0.5001025308121517) (4., 
  0.48731767677774623) (4.1000000000000005, 0.47692817103225704) (4.2,
   0.462229070923038) (4.3, 0.4540774587316336) (4.4, 
  0.44691172006107144) (4.5, 0.4323770717756109) (4.6000000000000005, 
  0.4242424282490129) (4.7, 0.41654540249343586) (4.800000000000001, 
  0.4092619261421842) (4.9, 0.3953075079114567) (5., 
  0.38857005583308996)};
\addplot[black,ultra thick,dotted,smooth]coordinates{(0., 1.7050233427052999) (0.1, 1.7057979764314362) (0.2, 
  1.7060292732747162) (0.30000000000000004, 1.7061321288883986) (0.4, 
  1.7062960897076451) (0.5, 1.7073804689674046) (0.6000000000000001, 
  1.7582113667807058) (0.7000000000000001, 1.7316433943044234) (0.8, 
  1.6869351481088763) (0.9, 1.6329468609614588) (1., 
  1.580797479550784) (1.1, 1.5406504390079176) (1.2000000000000002, 
  1.4269099446702442) (1.3, 1.33232579623711) (1.4000000000000001, 
  1.2582062199716109) (1.5, 1.1914771107759519) (1.6, 
  1.118854500137972) (1.7000000000000002, 1.0717113052664917) (1.8, 
  1.0189470729808356) (1.9000000000000001, 0.9726665081535413) (2., 
  0.9291583108221007) (2.1, 0.888868762682217) (2.2, 
  0.8522568221450054) (2.3000000000000003, 
  0.8183523618389401) (2.4000000000000004, 0.7858907066874553) (2.5, 
  0.7574159943688147) (2.6, 0.7291004976974955) (2.7, 
  0.7059132764082489) (2.8000000000000003, 
  0.6833076723273226) (2.9000000000000004, 0.6620182881336973) (3., 
  0.642241678289071) (3.1, 0.6187600679023476) (3.2, 
  0.6014328423849361) (3.3000000000000003, 
  0.5830090522602251) (3.4000000000000004, 0.5679424785492452) (3.5, 
  0.5538873718866221) (3.6, 0.5407733752528653) (3.7, 
  0.5285351507370499) (3.8000000000000003, 
  0.5155410085835889) (3.9000000000000004, 0.5047683513998938) (4., 
  0.49470867056214646) (4.1000000000000005, 0.47876491481658795) (4.2,
   0.4697100877599818) (4.3, 0.4563738228976351) (4.4, 
  0.4502196735333234) (4.5, 0.4426880153707414) (4.6000000000000005, 
  0.4263288263375817) (4.7, 0.41929815174992924) (4.800000000000001, 
  0.4117344450944335) (4.9, 0.3946896612046537) (5., 
  0.37967922837225765)};
\addplot[black,ultra thick,dashed,smooth]coordinates{(0., 1.4378251233492525) (0.1, 1.4380519967191552) (0.2, 
  1.4398336656885067) (0.30000000000000004, 1.4433898847270108) (0.4, 
  1.4479593109025986) (0.5, 1.4528927046929194) (0.6000000000000001, 
  1.4574777851095222) (0.7000000000000001, 1.4650348682084007) (0.8, 
  1.5791644114619625) (0.9, 1.56706510757469) (1., 
  1.561943385917421) (1.1, 1.5641059174147813) (1.2000000000000002, 
  1.4629009643738684) (1.3, 1.3577065521663938) (1.4000000000000001, 
  1.2760781698795) (1.5, 1.2101112484795649) (1.6, 
  1.1454518869889987) (1.7000000000000002, 1.0817262825797826) (1.8, 
  1.0267078743946907) (1.9000000000000001, 0.9768643593594154) (2., 
  0.9325721763857858) (2.1, 0.8921128189247073) (2.2, 
  0.8543708205671997) (2.3000000000000003, 
  0.8206568827372333) (2.4000000000000004, 0.7905076037381179) (2.5, 
  0.7614242211645471) (2.6, 0.7377179298309742) (2.7, 
  0.7179305465040134) (2.8000000000000003, 
  0.6976758950709667) (2.9000000000000004, 0.6794353613041644) (3., 
  0.6602058697620735) (3.1, 0.6453104907625092) (3.2, 
  0.6190194125564834) (3.3000000000000003, 
  0.6066712630746628) (3.4000000000000004, 0.5956244838144052) (3.5, 
  0.5703898708780251) (3.6, 0.5482152045471107) (3.7, 
  0.5392173444434227) (3.8000000000000003, 
  0.5221630327372969) (3.9000000000000004, 0.509933561590327) (4., 
  0.5032010022256639) (4.1000000000000005, 0.47496110356777166) (4.2, 
  0.46280764211656844) (4.3, 0.4626140520839201) (4.4, 
  0.4577992435532804) (4.5, 0.4535267975514582) (4.6000000000000005, 
  0.44975570160607836) (4.7, 0.43430269028103036) (4.800000000000001, 
  0.43100102098601806) (4.9, 0.42217225231619593) (5., 
  0.4194578039319957)};
\draw[black,thick,dotted] (axis cs:1.15,0.0) -- (axis cs:1.15,1.9);
\end{axis}
\end{tikzpicture}
}
\end{minipage}\hfill
\begin{minipage}[b]{0.5\columnwidth}
     \centering
        \resizebox{\linewidth}{!}{%
\centering
\pgfplotsset{every axis/.append style={font=\large,
extra description/.code={5
\node at (0.85,0.4) {\bf{\Large{(a)}}};
\node at (0.825,0.25) {\bf{\Large{(c)}}};
\node at (0.5,0.85){\bf{\Large{(d)}}};
\node at (0.5,0.125){1.15};
}}}
\begin{tikzpicture}[scale=1,line width=3pt]
\begin{axis}[
scale=1,
domain=0:2,xmax=2, xmin=0,ymax=2.85, ymin=0.0,
xlabel={\Large{$\bar{E}_{x}$}},
ylabel={\Large{$\chi^{0}\left(\bar{E}_{x}\right)$ (a.u.)}}
]
\addplot[blue,ultra thick,smooth] coordinates {(0., 0.9746741112973714) (0.05, 0.32110759703183683) (0.1, 
  0.3226872368096961) (0.15000000000000002, 0.32541360186348156) (0.2,
   0.32934766298041146) (0.25, 
  0.3345865543527462) (0.30000000000000004, 
  0.34127031375436984) (0.35000000000000003, 
  0.34958851277416886) (0.4, 0.5071011088255404) (0.45, 
  0.5360103599002242) (0.5, 0.5547588698511362) (0.55, 
  0.5762510196598082) (0.6000000000000001, 0.6423838504594443) (0.65, 
  0.6791562551678503) (0.7000000000000001, 0.7210033080348994) (0.75, 
  0.7681950554761566) (0.8, 0.823908685785057) (0.8500000000000001, 
  0.8842645696059063) (0.9, 0.949546996986306) (0.9500000000000001, 
  1.0796064135983885) (1., 1.1228617399270227) (1.05, 
  1.1382992950922206) (1.1, 1.3478459552109538) (1.1500000000000001, 
  1.8314196596056596) (1.2000000000000002, 1.8173044930109743) (1.25, 
  1.699124205380618) (1.3, 1.3507862387785108) (1.35, 
  1.1059422755459971) (1.4000000000000001, 
  1.0582036613387136) (1.4500000000000002, 1.0206168511979454) (1.5, 
  0.9900698928832266) (1.55, 0.9646371658079804) (1.6, 
  0.9431746586083171) (1.6500000000000001, 
  0.9248213374668798) (1.7000000000000002, 0.9089555053041292) (1.75, 
  0.8951140780334572) (1.8, 0.8829501285311704) (1.85, 
  0.8721772009583387) (1.9000000000000001, 
  0.8625782020429882) (1.9500000000000002, 0.8539791025251507) (2., 
  0.846238562774478)};
\addplot[black,ultra thick,dotted,smooth]coordinates{(0., 0.9728112229094186) (0.05, 0.3204031532583574) (0.1, 
  0.32191128113894407) (0.15000000000000002, 0.3245661220474084) (0.2,
   0.32841395419351804) (0.25, 
  0.33353618797900736) (0.30000000000000004, 
  0.34005202552416003) (0.35000000000000003, 
  0.34812288598275953) (0.4, 0.5052368704723993) (0.45, 
  0.5192584241603126) (0.5, 0.535682578827807) (0.55, 
  0.6091950927299065) (0.6000000000000001, 0.6418975734274682) (0.65, 
  0.6772272441916629) (0.7000000000000001, 0.7170187386001988) (0.75, 
  0.7615753697175702) (0.8, 0.814106390317549) (0.8500000000000001, 
  0.8712866698113498) (0.9, 0.9336091325324427) (0.9500000000000001, 
  1.281223429230364) (1., 1.2335221246944816) (1.05, 
  1.140701131781888) (1.1, 1.356444241366292) (1.1500000000000001, 
  2.0176513524189494) (1.2000000000000002, 1.917359837594253) (1.25, 
  1.6983284561349048) (1.3, 1.350869309654836) (1.35, 
  1.2723702915241142) (1.4000000000000001, 
  1.0590280285967473) (1.4500000000000002, 1.0211826389848346) (1.5, 
  0.9904568161933123) (1.55, 0.5284365061265884) (1.6, 
  0.5171249048407413) (1.6500000000000001, 
  0.5073308137669478) (1.7000000000000002, 0.49877553928318075) (1.75,
   0.49124573437053476) (1.8, 0.4845683520892843) (1.85, 
  0.4786175966951958) (1.9000000000000001, 
  0.4732873731582046) (1.9500000000000002, 0.4684909444860699) (2., 
  0.46415664971597437)};
\addplot[black,ultra thick,smooth]coordinates{(0., 0.9626618334090584) (0.05, 0.3171566363517817) (0.1, 
  0.3184886542887897) (0.15000000000000002, 0.320694967200927) (0.2, 
  0.3238271371906697) (0.25, 0.3279467894767711) (0.30000000000000004,
   0.33312498841225896) (0.35000000000000003, 
  0.4832129024974927) (0.4, 0.4930636362024092) (0.45, 
  0.5046735546184997) (0.5, 0.5181974753041148) (0.55, 
  0.534019910473212) (0.6000000000000001, 0.5520023639811728) (0.65, 
  0.572220393673039) (0.7000000000000001, 0.5947401882025387) (0.75, 
  0.7637273276226596) (0.8, 0.9754722521325524) (0.8500000000000001, 
  1.0419456482247422) (0.9, 1.1132918355327426) (0.9500000000000001, 
  1.1926356125220263) (1., 1.3451235563495032) (1.05, 
  1.4880850024631374) (1.1, 1.52096578894238) (1.1500000000000001, 
  2.48122469524139) (1.2000000000000002, 1.7397729961603885) (1.25, 
  0.9768310642296776) (1.3, 0.9088888759585771) (1.35, 
  0.8610215706469885) (1.4000000000000001, 
  0.8245064099424058) (1.4500000000000002, 0.7953898217018075) (1.5, 
  0.533603916276791) (1.55, 0.5204128386660078) (1.6, 
  0.5090992215694773) (1.6500000000000001, 
  0.49929140229375435) (1.7000000000000002, 
  0.49071289510622496) (1.75, 0.48315226485489865) (1.8, 
  0.4764397039843277) (1.85, 0.47045004389702627) (1.9000000000000001,
   0.4650785099742951) (1.9500000000000002, 0.46023942960113506) (2., 
  0.4558619491170608) (2.0500000000000003, 0.4518852184030674) (2.1, 
  0.448262070384403) (2.15, 0.4449507309183589) (2.2, 
  0.441915524162726) (2.25, 0.4391258171871035) (2.3000000000000003, 
  0.43655433145778205) (2.35, 0.4341793190275494) (2.4000000000000004,
   0.43198093128043147) (2.45, 0.4299417709225554) (2.5, 
  0.42804654620484905) (2.5500000000000003, 0.4262813480915517) (2.6, 
  0.424634836644626) (2.6500000000000004, 0.4230964786856765) (2.7, 
  0.4216568784018576) (2.75, 0.42030763697487455) (2.8000000000000003,
   0.4190410089215553) (2.85, 
  0.41785055748315353) (2.9000000000000004, 0.4167302370686912) (2.95,
   0.4156745823667828) (3., 0.4146786532500981) (3.0500000000000003, 
  0.4137378386129923) (3.1, 0.41284826005082725) (3.1500000000000004, 
  0.41200624652288625) (3.2, 0.41120836583560555) (3.25, 
  0.41045150761939825) (3.3000000000000003, 
  0.40973294514499586) (3.35, 
  0.40905026986942244) (3.4000000000000004, 0.4084010112424918) (3.45,
   0.4077829181095721) (3.5, 0.40719406679816667) (3.5500000000000003,
   0.4066325917475) (3.6, 0.4060968553322938) (3.6500000000000004, 
  0.40558529617998096) (3.7, 0.40509647203283433) (3.75, 
  0.4046290433968184) (3.8000000000000003, 0.40418173899542226) (3.85,
   0.40375344492054227) (3.9000000000000004, 
  0.40334309196688334) (3.95, 0.40294968315275137) (4., 
  0.40257229246072945) (4.05, 0.4022100362511157) (4.1000000000000005,
   0.40186212946858485) (4.15, 0.401527820413307) (4.2, 
  0.40120640652351824) (4.25, 0.40089722566947233) (4.3, 
  0.4005996494956084) (4.3500000000000005, 0.4003131143820225) (4.4, 
  0.4000370756872222) (4.45, 0.3997710250397569) (4.5, 
  0.3995144823553759) (4.55, 0.39926699616488953) (4.6000000000000005,
   0.3990281418279346) (4.65, 0.39879751924201634) (4.7, 
  0.3985747513630113) (4.75, 0.3983593936935102) (4.800000000000001, 
  0.3981512911628945) (4.8500000000000005, 0.39795012953539255) (4.9, 
  0.39775543335378494) (4.95, 0.39756701421061763) (5., 
  0.39738461074865805)};
\draw[black,thick,dotted] (axis cs:1.153,0.0) -- (axis cs:1.153,2.85);
\end{axis}
\end{tikzpicture}
}
\end{minipage}
\caption{(\emph{Left}) Spin-spin correlation as a function of an externally applied field. For $\bar{\Delta}_{0}=0.1$, (a) $J=0$, (b) $J=0.25$, (c) $J=0.5$ and (d) $J=1.0$. (\emph{Right}) DOS response. Case (b) (not shown) mostly overlaps on line (a).}\label{Dij2}
\end{figure}
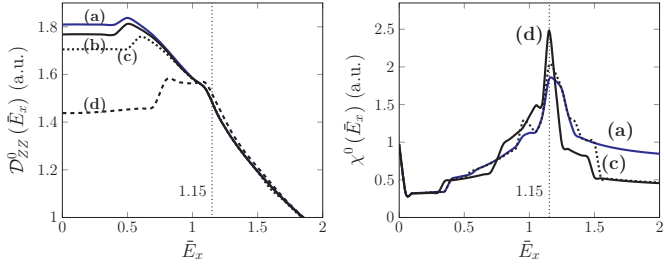

Figure (\ref{AAA}) represents the Fermi surface transformation under an applied field $\bar{E}_{x}$, low density regime and conserved number of particles $N_{-}$, calculated from Eq. (\ref{energy0}). $\bar{E}_{y}$ is taken as zero throughout all numerical calculations. Fermi surface topology at zero field corresponds to an annulus with radii $k_{F\sigma}$ \cite{Cap}. Applied field shifts the Fermi surface towards $k_{y}$ axis and it might eventually create a spin current on the same direction \cite{Jairo}.
Allowed states for non applied field lie into a circular Fermi disk with approximated radius $2k_{0}$. Compactness in the Fermi disk breaks into unconnected and asymmetric lobes at $\bar{E}_{x}=1.15$, phenomenon which is directly reflected in a strong peak on the DOS distribution (Figure \ref{Dij2}). Self consistent solutions of Eq.(\ref{szz1}) are shown in Fig. (\ref{muu})-(\emph{Left}). The normalized average spin polarization decreases with $\bar{E}_{x}$ and depends on the strength of $J$. Variations of $\langle \hat{S}_{Z}\rangle$ are more sensitive for small intensities of $\bar{E}_{x}$, and it tends to reach a weaker \emph{saturation} state for greater intensities. Strong Zeeman coupling \emph{favors} the effective spin alignment in the calculated range since $\langle\hat{S}_{Z}(J=1)\rangle<\langle \hat{S}_{Z}(J=0)\rangle<0$.  Fermi energy increases monotonically and is highly sensitive to the electric field intensity, although its rate of growing is lesser as the parameter $J$ gets stronger (Fig. \ref{muu})-(\emph{Right}).    

\begin{figure}
\centering
\pgfplotsset{every axis/.append style={
extra description/.code={5
\node at (0.15,0.275) {(b)};
\node at (0.6,0.8) {(a)};
\node at (0.8125,0.035) {(c)};
\node at (0.8125,0.625) {(e)};
\node at (0.575,0.08) {(d)};
}}}
\begin{tikzpicture}[scale=0.8125,line width=3pt]
\begin{axis}[scale only axis,
scale=0.8125,
domain=-1.25:0.20,xmax=0.20, xmin=-1.25,ymax=4.0, ymin=0,
axis y line*=left,
axis x line*=none,
xlabel={$\bar{\mu}$},
ylabel={$\chi^{0}\left(\bar{\mu}\right)$ (a.u.)}
]
\addplot[blue,thick,smooth] coordinates {(-0.7, 0.) (-0.69, 0.) (-0.6799999999999999, 0.) (-0.6699999999999999,
   0.) (-0.6599999999999999, 0.) (-0.6499999999999999, 
  0.) (-0.6399999999999999, 0.) (-0.6299999999999999, 0.) (-0.62, 
  0.) (-0.61, 0.) (-0.6, 0.) (-0.59, 0.) (-0.58, 0.) (-0.57, 
  0.) (-0.5599999999999999, 0.) (-0.5499999999999999, 
  0.) (-0.5399999999999999, 0.) (-0.5299999999999999, 0.) (-0.52, 
  0.) (-0.51, 0.) (-0.49999999999999994, 0.) (-0.49, 
  7.691146331908054) (-0.48, 5.191338325854772) (-0.47, 
  4.18299034911285) (-0.45999999999999996, 
  3.5997323865458912) (-0.44999999999999996, 
  3.207769131245473) (-0.43999999999999995, 
  2.9211378681082167) (-0.42999999999999994, 
  2.6997776414467585) (-0.41999999999999993, 
  2.5221601569547727) (-0.41, 
  2.3755458376806704) (-0.39999999999999997, 
  2.2518509299934695) (-0.38999999999999996, 
  2.145664596161109) (-0.37999999999999995, 
  2.0532110818439246) (-0.36999999999999994, 
  1.9717643740114543) (-0.35999999999999993, 
  1.8993003177547294) (-0.3499999999999999, 
  1.834280555766041) (-0.33999999999999997, 
  1.7755119359214477) (-0.32999999999999996, 
  1.7220528440740408) (-0.31999999999999995, 
  1.6731491461281052) (-0.30999999999999994, 
  1.6281882835992647) (-0.29999999999999993, 
  1.5866679882702102) (-0.2899999999999999, 
  1.5481702762657052) (-0.27999999999999997, 
  1.5123461049592644) (-0.26999999999999996, 
  1.4788976984753583) (-0.25999999999999995, 
  1.4475755506401673) (-0.24999999999999994, 
  1.4181628166126359) (-0.23999999999999994, 
  1.3904730737878372) (-0.22999999999999993, 
  1.3332395788671603) (-0.21999999999999997, 
  1.3096162046393496) (-0.20999999999999996, 
  1.2872280694782612) (-0.19999999999999996, 
  1.2659723406108647) (-0.18999999999999995, 
  1.2457572435333397) (-0.17999999999999994, 
  1.2265013566561846) (-0.16999999999999993, 
  1.2081320169594543) (-0.15999999999999992, 
  1.1905841499129755) (-0.1499999999999999, 
  1.1737996018962789) (-0.1399999999999999, 
  1.1577255435779272) (-0.1299999999999999, 
  1.1423147188563447) (-0.12, 
  1.1275242620358434) (-0.10999999999999999, 
  1.0711080209631576) (-0.09999999999999998, 
  1.061996839335481) (-0.08999999999999997, 
  1.053231157230395) (-0.07999999999999996, 
  1.0034565937416973) (-0.06999999999999995, 
  1.0013234808450724) (-0.05999999999999994, 
  0.9862049357237379) (-0.04999999999999993, 
  0.9843957018191749) (-0.039999999999999925, 
  0.9826441121289816) (-0.029999999999999916, 
  0.9809472952620816) (-0.019999999999999907, 
  0.979302456104293) (-0.009999999999999898, 
  0.9777070941878467) (0.000001, 
  0.9761588446169368) (0.010000000000000009, 
  0.9683746430653399) (0.020000000000000018, 
  0.9608155173920764) (0.030000000000000027, 
  0.9534707392869439) (0.040000000000000036, 
  0.9750798670412774) (0.050000000000000044, 
  0.9672028174954705) (0.06000000000000005, 
  0.9605874078870922) (0.07000000000000006, 
  0.9585455401795044) (0.08000000000000007, 
  0.9486887943829437) (0.09000000000000008, 
  0.9459125704280229) (0.10000000000000009, 
  0.9612256910750033) (0.1100000000000001, 
  0.9917059530351245) (0.1200000000000001, 
  0.9931145427079586) (0.13000000000000012, 0.9908954455627421) (0.14,
   0.9958514876857522) (0.15000000000000002, 
  0.997180409719757) (0.16000000000000003, 
  0.9921533422176888) (0.17000000000000004, 
  0.9995760893877956) (0.18000000000000005, 
  1.0010136793466888) (0.19000000000000006, 
  1.0022423760822814) (0.20000000000000007, 0.9972547162773449)};
\addplot[black,ultra thick,smooth,dotted] coordinates {(-1.24, 0.) (-1.23, 0.655021764956693) (-1.22, 
  0.6566244075661487) (-1.21, 0.6575952883407653) (-1.2, 
  0.6585846778060928) (-1.19, 0.6595932334666419) (-1.18, 
  0.6606215846926407) (-1.17, 0.6616704036287481) (-1.16, 
  0.6627403858737136) (-1.15, 0.6638322605117355) (-1.14, 
  0.6649467951582866) (-1.13, 
  0.6660848000018139) (-1.1199999999999999, 
  0.6672471229493497) (-1.1099999999999999, 
  0.6684346603383409) (-1.0999999999999999, 
  0.6696483386845584) (-1.0899999999999999, 0.670889153006299) (-1.08,
   1.3427421694572084) (-1.0699999999999998, 
  1.3469128499187055) (-1.06, 1.3495704003889293) (-1.05, 
  1.3522912453241065) (-1.04, 1.3550780648403915) (-1.03, 
  1.3579336129121227) (-1.02, 1.360860816952874) (-1.01, 
  1.3638627922067852) (-1., 1.3669428398081662) (-0.99, 
  1.3701044722205074) (-0.98, 1.3733514448564104) (-0.97, 
  1.3766877479229314) (-0.96, 1.3801176537200128) (-0.95, 
  1.3836456964536337) (-0.94, 
  1.3872768250755163) (-0.9299999999999999, 
  1.3910162328982267) (-0.9199999999999999, 
  1.3948695647896443) (-0.9099999999999999, 
  1.3988428743317258) (-0.8999999999999999, 
  1.4029427024298418) (-0.8899999999999999, 
  1.4071760917830602) (-0.8799999999999999, 
  1.4115506858011238) (-0.8699999999999999, 
  1.4160747035971792) (-0.8599999999999999, 
  1.4207571400670969) (-0.8499999999999999, 
  1.4256077735773147) (-0.84, 1.4306372324569847) (-0.83, 
  1.4358571735920136) (-0.82, 1.44128034225856) (-0.8099999999999999, 
  1.446920774484908) (-0.7999999999999999, 
  1.4527939125609042) (-0.7899999999999999, 
  1.4589168466447362) (-0.78, 1.2353821649421481) (-0.77, 
  1.2417094175085013) (-0.76, 1.2483262340040737) (-0.75, 
  1.2552572673239952) (-0.74, 1.2625302641553087) (-0.73, 
  1.270176635903823) (-0.72, 1.278232114304059) (-0.71, 
  1.2867375521599473) (-0.7, 1.2957400965394672) (-0.69, 
  1.305294188518313) (-0.6799999999999999, 
  1.315463433016313) (-0.6699999999999999, 
  1.3263229412276543) (-0.6599999999999999, 
  1.337961419452283) (-0.6499999999999999, 
  1.3504855756661804) (-0.6399999999999999, 
  1.3640246937244827) (-0.6299999999999999, 
  1.3787375918828808) (-0.62, 1.3948220546646792) (-0.61, 
  1.4125289935471466) (-0.6, 1.4321820955963167) (-0.59, 
  1.4542097121818234) (-0.58, 1.4791939210093883) (-0.57, 
  1.5079534413696325) (-0.5599999999999999, 
  1.7836463242554665) (-0.5499999999999999, 
  1.826550776699265) (-0.5399999999999999, 
  1.8797646934272214) (-0.5299999999999999, 
  1.9415877887389201) (-0.5199999999999999, 
  2.04014697616953) (-0.5099999999999999, 
  2.10115980038701) (-0.4999999999999999, 2.7763569553716763) (-0.49, 
  1.7093384789836557) (-0.48, 1.5253228466149193) (-0.47, 
  1.406282623301242) (-0.45999999999999996, 
  1.3162353776867974) (-0.44999999999999996, 
  1.2429807421191748) (-0.43999999999999995, 
  1.1808054962564725) (-0.42999999999999994, 
  1.12654351495012) (-0.41999999999999993, 
  1.0782457185779444) (-0.4099999999999999, 
  1.0346209309710226) (-0.3999999999999999, 
  0.9947664125129426) (-0.3899999999999999, 
  0.9580243414437288) (-0.3799999999999999, 
  0.816243947547566) (-0.3699999999999999, 0.7902125634658717) (-0.36,
   0.7657153691219467) (-0.35, 
  0.7425586936483539) (-0.33999999999999997, 
  0.6291056748197095) (-0.32999999999999996, 
  0.6130374895759297) (-0.31999999999999995, 
  0.5976468004829916) (-0.30999999999999994, 
  0.5828639419274437) (-0.29999999999999993, 
  0.5686285771922861) (-0.2899999999999999, 
  0.5548879151671728) (-0.2799999999999999, 
  0.5415951913123604) (-0.2699999999999999, 
  0.5287086936640374) (-0.2599999999999999, 
  0.6252359121759966) (-0.2499999999999999, 
  0.6159766360570125) (-0.24, 
  0.607113387618495) (-0.22999999999999998, 
  0.5986201816389635) (-0.21999999999999997, 
  0.5904738395185247) (-0.20999999999999996, 
  0.5826533248060827) (-0.19999999999999996, 
  0.5751399161475003) (-0.18999999999999995, 
  0.5679162238448114) (-0.17999999999999994, 
  0.5609671040282391) (-0.16999999999999993, 
  0.6142985551284185) (-0.15999999999999992, 
  0.6087748673236648) (-0.1499999999999999, 
  0.6034605909261292) (-0.1399999999999999, 
  0.5983471219008708) (-0.1299999999999999, 
  0.5934276535224573) (-0.11999999999999988, 
  0.5886976019085053) (-0.10999999999999988, 
  0.5841557224602779) (-0.09999999999999987, 
  0.5798062963144734) (-0.08999999999999986, 
  0.5756648141907251) (-0.07999999999999985, 
  0.5717761994073356) (-0.06999999999999984, 
  0.568313203002843) (-0.05999999999999983, 
  0.5654493210426572) (-0.04999999999999982, 
  0.5609536116450689) (-0.040000000000000036, 
  0.5565901127201501) (-0.030000000000000027, 
  0.5523567089640357) (-0.020000000000000018, 
  0.5482556079785443) (-0.010000000000000009, 0.5152949528372949) (0.,
   0.5121856402508634) (0.010000000000000009, 
  0.5091424016713352) (0.020000000000000018, 
  0.5061626349145765) (0.030000000000000027, 
  0.5032440857848511) (0.040000000000000036, 
  0.5003844123525484) (0.050000000000000044, 
  0.49758162878721196) (0.06000000000000005, 
  0.49483372934588493) (0.07000000000000006, 
  0.4907978056238513) (0.08000000000000007, 
  0.4866840778320029) (0.09000000000000008, 
  0.4830720963049325) (0.10000000000000009, 
  0.47966029934977894) (0.1100000000000001, 
  0.4763773220227877) (0.1200000000000001, 
  0.473192679542187) (0.13000000000000012, 
  0.4700898990518798) (0.14000000000000012, 
  0.4670587550306012) (0.15000000000000013, 
  0.46409225368905405) (0.16000000000000014, 
  0.4611852818875483) (0.17000000000000015, 
  0.4583339009923384) (0.18000000000000016, 
  0.5349807979230565) (0.19000000000000017, 
  0.5325367867339862) (0.20000000000000018, 0.45008389878748256)};
\addplot[Mygreen,ultra thick,smooth] coordinates {(-0.53, 0.0) (-0.28, 0.0) (-0.27, 0.0) (-0.26, 0.0) (-0.25, 
  0.0) (-0.24000000000000005, 0.0) (-0.23000000000000004, 
  0.0) (-0.22000000000000003, 0.0) (-0.21000000000000002, 0.0) (-0.2, 
  0.0) (-0.19, 0.0) (-0.18, 0.0) (-0.17000000000000004, 
  0.0) (-0.16000000000000003, 0.0) (-0.15000000000000002, 0.0) (-0.14,
   0.0) (-0.13, 0.0) (-0.12, 0.0) (-0.11000000000000004, 
  0.0) (-0.10000000000000003, 0.0) (-0.09000000000000002, 
  0.0) (-0.08000000000000002, 0.0) (-0.07, 0.0) (-0.06, 
  0.000011597510924257374) (-0.050000000000000044, 
  0.00021364980915860107) (-0.040000000000000036, 
  0.0024558108650497975) (-0.030000000000000027, 
  0.017794798100461806) (-0.020000000000000018, 
  0.08258208370139962) (-0.010000000000000009, 
  0.2517375641481507) (0., 0.5250000000000007) (0.010000000000000009, 
  0.7982624358518507) (0.020000000000000018, 
  0.967417916298602) (0.030000000000000027, 
  1.0322052018995396) (0.040000000000000036, 
  1.047544189134952) (0.04999999999999993, 
  1.0497863501908429) (0.05999999999999994, 
  1.0499884024890775) (0.06999999999999995, 
  1.0499996098733562) (0.07999999999999996, 
  1.049999991905941) (0.08999999999999997, 
  1.0499999998967702) (0.09999999999999998, 
  1.0499999999991942) (0.10999999999999999, 1.049999999999998) (0.12, 
  1.0500000000000014) (0.13, 1.0500000000000014) (0.14, 
  1.0500000000000016) (0.15000000000000002, 
  1.050000000000002) (0.16000000000000003, 
  1.0500000000000018) (0.17000000000000004, 
  1.0500000000000016) (0.17999999999999994, 
  1.0500000000000014) (0.18999999999999995, 
  1.050000000000002) (0.19999999999999996, 1.0500000000000014)};
\draw[black,thick,dotted] (axis cs:-0.5,0.0) -- (axis cs:-0.5,6.0);
\draw[black, thick,dashed] (axis cs:-1.2,1.0) -- (axis cs:0.2,1.0);
\end{axis}
\begin{axis}[scale only axis,
scale=0.8125,
domain=-1.25:0.20,xmax=0.20, xmin=-1.25,ymax=4.0, ymin=0,
axis y line*=right,
ylabel={$\mathcal{D}_{ZZ}^{0}\left(\bar{\mu}\right)$ (a.u.)}
]
\addplot[red, thick,smooth] coordinates {(-0.55, 0.) (-0.54, 0.0) (-0.53, 0.0) (-0.52, 0.0) (-0.51, 0.0) (-0.5,
   0.0) (-0.49000000000000005, 
  0.2723187055649257) (-0.48000000000000004, 
  0.39284325856007535) (-0.47000000000000003, 
  0.484120213237611) (-0.4600000000000001, 
  0.5607082372837131) (-0.45000000000000007, 
  0.6280178798768099) (-0.44000000000000006, 
  0.6887777438065343) (-0.43000000000000005, 
  0.7445943720096281) (-0.42000000000000004, 
  0.7965083428661691) (-0.41000000000000003, 
  0.8452389619674873) (-0.4, 0.8913095234484294) (-0.39, 
  0.9351121850011402) (-0.38, 
  0.9769526923367611) (-0.37000000000000005, 
  1.0170770729097487) (-0.36000000000000004, 
  1.0556734880701206) (-0.35000000000000003, 
  1.0929193358427811) (-0.3400000000000001, 
  1.1289532882672757) (-0.33000000000000007, 
  1.1638178516347402) (-0.32000000000000006, 
  1.1976822535369531) (-0.31000000000000005, 
  1.2306272372821205) (-0.30000000000000004, 
  1.2627129334077483) (-0.29000000000000004, 
  1.2940031816267932) (-0.28, 1.3245541381748085) (-0.27, 
  1.3544185357834848) (-0.26000000000000006, 
  1.3836364581166773) (-0.25000000000000006, 
  1.4122497757387384) (-0.24000000000000005, 
  1.4402968417545998) (-0.23000000000000004, 
  1.4678061567346004) (-0.22000000000000003, 
  1.4948170914072962) (-0.21000000000000002, 1.521335505215597) (-0.2,
   1.5474091125548424) (-0.19000000000000006, 
  1.5730462112666368) (-0.18000000000000005, 
  1.5982715502626768) (-0.17000000000000004, 
  1.623112060286144) (-0.16000000000000003, 
  1.6475709886619345) (-0.15000000000000002, 
  1.6716837537842366) (-0.14, 
  1.6954411333264463) (-0.13000000000000006, 
  1.7188895856413815) (-0.12000000000000005, 
  1.7420015707442902) (-0.11000000000000004, 
  1.7648104009186072) (-0.10000000000000003, 
  1.7873296545629775) (-0.09000000000000002, 
  1.8095680689742557) (-0.08000000000000002, 
  1.831536568409198) (-0.07000000000000006, 
  1.853291532763781) (-0.06000000000000005, 
  1.8749880755868074) (-0.050000000000000044, 
  1.8959617013540633) (-0.040000000000000036, 
  1.9170005669635604) (-0.030000000000000027, 
  1.9377518161993612) (-0.020000000000000018, 
  1.9582836387097975) (-0.010000000000000009, 1.9786023736831775) (0.,
   1.9982282218455738) (0.010000000000000009, 
  1.999618526453696) (0.020000000000000018, 
  1.9995147946454739) (0.029999999999999916, 
  1.9995208519065844) (0.039999999999999925, 
  1.9994976828341393) (0.04999999999999993, 
  1.999503787174741) (0.05999999999999994, 
  1.9994683066722467) (0.06999999999999995, 
  1.9994744676867702) (0.07999999999999996, 
  1.9994804116640914) (0.08999999999999997, 
  1.999486252459454) (0.09999999999999998, 
  1.9994918813103462) (0.10999999999999999, 1.9995242812771066) (0.12,
   1.9996249005702098) (0.13, 1.999628741977749) (0.14, 
  1.9996325031664315) (0.15000000000000002, 
  1.9989247480165202) (0.15999999999999992, 
  1.999517274578364) (0.16999999999999993, 
  1.9996419507527405) (0.17999999999999994, 
  1.9996466000350994) (0.18999999999999995, 
  1.9996499350796813) (0.19999999999999996, 1.9992773307064975)};
\addplot[red, ultra thick,smooth,dotted] coordinates {(-1.2, 0.03466222821915727) (-1.19, 0.04336410667914071) (-1.18, 
  0.052098093794069315) (-1.17, 0.06086471657561575) (-1.16, 
  0.06966451947352095) (-1.15, 0.07849806263327329) (-1.14, 
  0.08736592541750299) (-1.13, 
  0.09626870546836548) (-1.1199999999999999, 
  0.10520702169358773) (-1.1099999999999999, 
  0.11418150438399488) (-1.0999999999999999, 
  0.12319282175855208) (-1.0899999999999999, 
  0.13224164462615448) (-1.08, 
  0.14953038602738833) (-1.0699999999999998, 
  0.1674915266892231) (-1.06, 0.18552345798456849) (-1.05, 
  0.20363342603589993) (-1.04, 0.22182300265149774) (-1.03, 
  0.2400938291072402) (-1.02, 0.25844760422321283) (-1.01, 
  0.2768861024454118) (-1., 0.29531701474540767) (-0.99, 
  0.3140247117556485) (-0.98, 0.33272874539047326) (-0.97, 
  0.3515254106751531) (-0.96, 0.37041671297263074) (-0.95, 
  0.3894051343571757) (-0.94, 
  0.40830614735288967) (-0.9299999999999999, 
  0.42768269952741267) (-0.9199999999999999, 
  0.4469770025379274) (-0.9099999999999999, 
  0.46637859548494776) (-0.8999999999999999, 
  0.4858867420462309) (-0.8899999999999999, 
  0.5055154002224443) (-0.8799999999999999, 
  0.5252568480397423) (-0.8699999999999999, 
  0.5451181212013715) (-0.8599999999999999, 
  0.5651027867260041) (-0.8499999999999999, 
  0.5852147092562249) (-0.84, 0.6053815205276721) (-0.83, 
  0.6258361453549532) (-0.82, 
  0.6463545733265447) (-0.8099999999999999, 
  0.6670177894059248) (-0.7999999999999999, 
  0.6878309401536331) (-0.7899999999999999, 
  0.7087995398590259) (-0.78, 0.7299295637093526) (-0.77, 
  0.7512270810432873) (-0.76, 0.7726992810460249) (-0.75, 
  0.7943533773834373) (-0.74, 0.8160889652594892) (-0.73, 
  0.8381940113281285) (-0.72, 0.8604451439001691) (-0.71, 
  0.8829148862653936) (-0.7, 0.9056148213746401) (-0.69, 
  0.9285572798525539) (-0.6799999999999999, 
  0.9516675810154842) (-0.6699999999999999, 
  0.975276442856227) (-0.6599999999999999, 
  0.9989022703704702) (-0.6499999999999999, 
  1.0229750263304063) (-0.6399999999999999, 
  1.0474210369658383) (-0.6299999999999999, 
  1.0721588728457518) (-0.62, 1.0973142915525385) (-0.61, 
  1.1228953441032732) (-0.6, 1.1489457168805617) (-0.59, 
  1.1754613878545244) (-0.58, 1.2025890474012064) (-0.57, 
  1.2303883818495986) (-0.5599999999999999, 
  1.2589504546244565) (-0.5499999999999999, 
  1.2885265049726409) (-0.5399999999999999, 
  1.3190127916264958) (-0.5299999999999999, 
  1.3511342225087206) (-0.5199999999999999, 
  1.385286822253503) (-0.5099999999999999, 
  1.4228638869440513) (-0.4999999999999999, 1.472569524920192) (-0.49,
   1.5105211624723869) (-0.48, 1.5443604811109684) (-0.47, 
  1.5760766909791837) (-0.45999999999999996, 
  1.6061223482115194) (-0.44999999999999996, 
  1.6348459619410944) (-0.43999999999999995, 
  1.662547392946034) (-0.42999999999999994, 
  1.6892882717775621) (-0.41999999999999993, 
  1.715217950931343) (-0.4099999999999999, 
  1.7404047586920122) (-0.3999999999999999, 
  1.7649349094770546) (-0.3899999999999999, 
  1.7888848846479457) (-0.3799999999999999, 
  1.8122244773140614) (-0.3699999999999999, 
  1.8350392913049092) (-0.36, 1.8574829589539767) (-0.35, 
  1.8794236838267755) (-0.33999999999999997, 
  1.9009510021598812) (-0.32999999999999996, 
  1.922105650298252) (-0.31999999999999995, 
  1.942806402202973) (-0.30999999999999994, 
  1.9630978429656267) (-0.29999999999999993, 
  1.9830257272251455) (-0.2899999999999999, 
  2.0025784867459424) (-0.2799999999999999, 
  2.0217593986769007) (-0.2699999999999999, 
  2.0406519043674525) (-0.2599999999999999, 
  2.0589901606280097) (-0.2499999999999999, 
  2.0770436397343115) (-0.24, 
  2.0947400056066576) (-0.22999999999999998, 
  2.1120442926675773) (-0.21999999999999997, 
  2.126032511854436) (-0.20999999999999996, 
  2.145469188093019) (-0.19999999999999996, 
  2.1615468828229787) (-0.18999999999999995, 
  2.1771728519736695) (-0.17999999999999994, 
  2.192364770270713) (-0.16999999999999993, 
  2.2070130969014805) (-0.15999999999999992, 
  2.221109537064281) (-0.1499999999999999, 
  2.2346954113876767) (-0.1399999999999999, 
  2.24767182463548) (-0.1299999999999999, 
  2.26005116557006) (-0.11999999999999988, 
  2.271760588010228) (-0.10999999999999988, 
  2.282787372480067) (-0.09999999999999987, 
  2.2931069915549194) (-0.08999999999999986, 
  2.302702591594809) (-0.07999999999999985, 
  2.31156625074147) (-0.06999999999999984, 
  2.319701404100809) (-0.05999999999999983, 
  2.309897831067772) (-0.04999999999999982, 
  2.3177261535823948) (-0.040000000000000036, 
  2.3259624187668666) (-0.030000000000000027, 
  2.334139542481306) (-0.020000000000000018, 
  2.342257628702042) (-0.010000000000000009, 2.350280038623334) (0., 
  2.3583099979914586) (0.010000000000000009, 
  2.366261661481177) (0.020000000000000018, 
  2.374160850794481) (0.030000000000000027, 
  2.382006519263724) (0.040000000000000036, 
  2.389802822594837) (0.050000000000000044, 
  2.3975312772619466) (0.06000000000000005, 
  2.4052629070570575) (0.07000000000000006, 
  2.445692664876646) (0.08000000000000007, 
  2.456114155814068) (0.09000000000000008, 
  2.464335031501332) (0.10000000000000009, 
  2.471741114772612) (0.1100000000000001, 
  2.478344929967444) (0.1200000000000001, 
  2.4842065016819173) (0.13000000000000012, 
  2.4894336736988) (0.14000000000000012, 
  2.4940702927413763) (0.15000000000000013, 
  2.498180317940655) (0.16000000000000014, 
  2.5018116579247773) (0.17000000000000015, 
  2.5050457692525763) (0.18000000000000016, 
  2.507924106659743) (0.19000000000000017, 
  2.510537984900004) (0.20000000000000018, 2.512811815276188)};
\end{axis}
\end{tikzpicture}
\caption{Real part of bare susceptibility response $\chi^{0}\left(\bar{\mu}\right)$ for non interacting spins ($J=0$) at zero temperature. (a) $\bar{\Delta}_{0}=0$, $\bar{E}_{x}=0$ (b) $\bar{\Delta}_{0}=0.1$, $\bar{E}_{x}=1.15$. (c) Density of states line for a free electron system at zero field calculated from $\chi^{0}$. (d) $\mathcal{D}_{ZZ}^{0}\left(\bar{\mu}\right)$ [right axis] at $\bar{E}_{x}=\bar{\Delta}_{0}=0$, (e) $\bar{E}_{x}=1.15$, $\bar{\Delta}_{0}=0.1$.}\label{X0}
\end{figure}
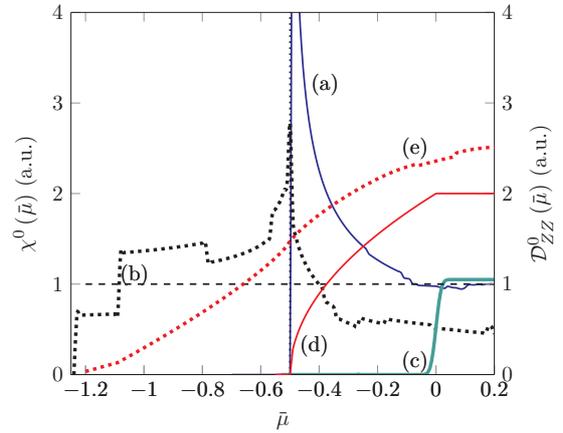
 Figure (\ref{X0}) shows the bare susceptibility at zero temperature as a function of the Fermi energy. This quantity recasts the density of states for the 2D case \emph{without} interaction, exhibiting a strong peak at $\bar{\mu}=-1/2$ [line (a)] \cite{Cap}. Line (b) describes the DOS distribution for $\bar{E}_{x}=1.15$. Peaks on the DOS are also centered around $\bar{\mu}=-1/2$, with cut off points and intensity depending on the electric field magnitude. Line (c) describes the DOS for a free electron system, constant for positive Fermi energy and zero otherwise. The spin-spin correlation function $\mathcal{D}_{ZZ}^{0}\left(\bar{\mu}\right)$ [line (d)] decreases to zero in the interval $-1/2<\bar{\mu}<0$ at zero field and no gap. Line (e) takes into account the gap and the applied field simultaneously, showing a higher cut off value for $\bar{\mu}$. No restriction for the total number of particles has been imposed here, nevertheless, this results shall be considered as the benchmark for comparative purposes and further calculations. Correlation function and DOS response for interacting spins at zero temperature are shown in Figure (\ref{Dij2}). Real part of $\mathcal{D}_{ZZ}^{0}\left(\bar{E}_{x}\right)$ behaves uniformly for $\bar{E}_{x}< 1.15$ and rapidly decays for $\bar{E}_{x}>1.15$ at $J=1$. The coupling effect $J$ on the spin-spin correlation $\mathcal{D}_{ZZ}^{0}(\bar{E}_{x})$ is important for smaller values of $\bar{E}_{x}$ ($\bar{E}_{x}<1.15$ in this case) and it is insignificant for stronger intensities, providing a signature for which $\langle \hat{S}_{Z}\rangle$ enters into a state of saturation dominated by the electric field over the band gap energy effect $\bar{\Delta}_{0}$; i.e., the $Z$-spin susceptibility $\chi_{ZZ}^{0}(\bar{E}_{x}\sim 0)$ tends to decrease in magnitude as the $J$ parameter augments in the difference associate to the intrinsic Rashba field $\gamma B_{\Sigma Z}=-\Delta_{0}-J\langle\hat{S}_{Z}\rangle$, since $\langle\hat{S}_{Z}\rangle<0$. The function $\chi^{0}(\bar{E}_{x})$ exhibits a strong maximum associated to the geometric breach for the Fermi surface at $\bar{E}_{x}=1.15$ and $J=1$.  
\section{Concluding Remarks}
The classical two bands Rashba-type Hamiltonian resembles the Dirac electron equation and it may recast a Zeeman-like behavior for expanded spin basis and externally applied electric field. By using standard finite temperature formalism, we have computed the average spin polarization perpendicular to the confinement layer, the DOS and the generic spin-spin correlation functions in the long wavelength approach. The value $\bar{E}_{x}\approx 1.15$ yields in the order of $\alpha/e\ell^{2}$, ($e-$electron charge) providing an estimation for the lateral gate voltage in terms of the length scale $\ell$. The bare \emph{polarization propagator} $\chi^{0}$ recalls strong DOS fluctuations for an applied electric field, indicating a close relationship between its geometric distribution over the Fermi surface and the outbreak limit for conserved number of carriers. This model suggests the formation of a magnetic state via Zeeman-Rashba-type field $B_{\Sigma Z}$ at zero temperature and zero electric field due to presence of the energy band gap, or even for gapless configurations, where exchange spin spin-average interaction is taken into account. Temperature effects, different average spin directions associated to $\mathbf{B}_{\Sigma}$, as well as the nature of the spin-spin interaction have to be reconsidered beyond the purely \emph{on-site and collinear} coupling, possibly a RKKY-type interaction \cite{Bruno}, issues that shall be addressed on further investigations. 
\section*{Acknowledgements}
The author acknowledges availability and technical support at \emph{Bloque}-W Computer Lab, UN Manizales.  

 \end{document}